\documentclass[aip,pop,amsmath,reprint]{revtex4-1}
\usepackage{graphicx}
\usepackage{amsmath}
\usepackage{epstopdf}
\usepackage{float}
\usepackage{wrapfig}
\usepackage{afterpage}
\usepackage{hyperref}
\hypersetup{
	colorlinks=true,
	linkcolor=blue,
	urlcolor=blue,
	citecolor=blue,
}

\begin{document}
\title{Electron sheath evolution controlled by a magnetic field in modified hollow cathode glow discharge}

\author{R. Rane}
\email{ramu@ipr.res.in}
\affiliation{Institute for Plasma Research, Gandhinagar, India}
\affiliation{Homi Bhabha National Institute, Anushakti Nagar, Mumbai, India}
\author{S. Chauhan}
\affiliation{Institute for Plasma Research, Gandhinagar, India}
\author{P. Bharathi}
\affiliation{Institute for Plasma Research, Gandhinagar, India}
\author{K. Nigam}%
\affiliation{Institute for Plasma Research, Gandhinagar, India}
\author{P. Bandyopadhyay}%
\affiliation{Institute for Plasma Research, Gandhinagar, India}
\affiliation{Homi Bhabha National Institute, Anushakti Nagar, Mumbai, India}
\author{S. Mukherjee}%
\affiliation{Institute for Plasma Research, Gandhinagar, India}
\affiliation{Homi Bhabha National Institute, Anushakti Nagar, Mumbai, India}
\begin{abstract}
The electron sheath formation in a DC magnetised plasma of modified hollow cathode source is studied. The discharge consists of two plane parallel cathodes and a small cubical anode placed off axis at the center. The argon plasma is produced and the properties of the plasma in response to the sheath formation near the anode are studied using electrical and optical diagnostics. In particular, the effect of pressure, magnetic field on discharge parameters such as discharge current, plasma potential, plasma density and electron temperature is studied. The discharge showed an onset of anode glow at a critical applied magnetic field indicating the formation of electron sheath and a double layer. The discharge current initially decreases; however it starts to rise again as the anode spot appears on the anode. The critical magnetic field at which anode glow formation takes place is dependent upon operating pressure and discharge voltage. The transition from ion sheath to electron sheath is investigated in detail by Langmuir probe and spectroscopy diagnostics. The plasma potential near  the  anode  decreases  during the transition from ion sheath to electron sheath. The plasma potential locks to the ionization potential of argon gas when anode spot is completely formed. A systematic study showed that during the transition, the electron temperature increases and plasma density decreases in the bulk plasma. The spectroscopy of the discharge showed presence of strong atomic and ionic lines of argon. The intensity of these spectral lines showed a dip during the transition between two sheaths. After the formation of the anode spot, oscillations of the order of 5-20 kHz are observed in the discharge  current and  floating potential due to the enhanced  ionisation and excitation processes  in the electron  sheath. The reason of the electron sheath formation at particular magnetic field is attributed to the reduction of the electron flux reaching to the anode in the direction perpendicular to the magnetic field.
\end{abstract}
\maketitle
\section{Introduction}
Hollow cathode discharges are relatively efficient ion, electron, and plasma sources where dense plasma can be produced in an electrostatic cavity formed by cathodes. Ion sheaths at the cathodes act as reflectors and confine electrons in the bulk plasma. The electrostatic confinement of electrons leads to an efficient ionization of the working gas. With an additional applied magnetic field along the axial direction, the hollow cathode source can be modified to a Penning type source\citep{Toader2004}. With this modification, electrons trying to escape towards the anode are now tied to the magnetic field lines. Thus, electrons move in cycloid paths and trapped in axial potential well. Because of increased path lengths the discharge becomes self-sustaining at pressures far below the operating pressures of a normal glow discharge. Interestingly, it is also possible to modify these discharges as electron sources by creating a global non-ambipolar flow\citep{Longmier2006} where all the electrons are lost via electron sheaths and all ions are lost to cathode via ion sheaths. 
Plasma sheaths are ubiquitous, non-neutral regions which are formed at plasma material interface to balance electron and ion losses. In low temperature plasma applications, the sheaths are important for providing directed ion energy for the surface modification. Similarly in high temperature plasmas, sheaths determine the wall heating, erosion etc. This important structure in plasma physics has been studied extensively \citep{scott2013, Brett2016, Franklin2003, Stenzel2011, Hershkowitz2005}. There are two varieties of sheath i.e. ion sheath and electron sheath. The ion rich sheath is a thin positive space charge layer that limits the electron losses to the boundary, to maintain the quasi-neutrality. This type of sheaths are common due to the large mobility of the electrons. In most of the devices, in which the electron losses have been studied, the larger part of electrons were lost through the ion sheaths. The sheath near the small electrode can be electron sheath when the electrode is biased positive with respect to the plasma potential. This sheath is formed near the small electrode (or anode) when the ratio of its area (i.e. electron collecting area, $A_e$)to the ion collecting area (i.e. $A_i$) satisfies $A_e/A_i < \sqrt{2.3m_e/m_i}$\citep{baalrud2009equilibrium}. The electron sheaths are generally observed around Langmuir probe \citep{Langmuir1926}, plasma contactor\citep{song1991anode}, magnetically constricted electrode\citep{chauhan2016droplet} etc. The electron sheath has been also characterised at low density and low temperature plasmas\citep{Bailung2017}. The transition of electron sheath and its different states like anode glow and anode spot has been studied in detail\citep{conde2006transition, baalrud2009equilibrium}. Recently, Scheiner \textit{et. al.} developed a 1D model for studying the electron sheath and pre-sheath using particle in cell (PIC) simulation\cite{baalrud-scheiner-theory-ES}. Yee et.al. have also reported the experimental observation of long electron sheath region extending into the plasma\citep{EsheathYEE}. The evolution of anode sheath/glow into double layer in magnetised plasma has been studied experimentally by biasing auxiliary electrode above the plasma potential\cite{torven1979dl, tang2003anode, Andersson1981, steven1987}. The sheath formed between magnetized plasma and particle absorbing wall (e.g. Anode) is  studied in past\citep{Holland1993, chodura1982}. Although electron sheaths have been studied extensively, few experiments are reported on the evolution of electron sheath on the discharge electrode itself  due to the variation of magnetic field. This type study is important considering its effect in sources which are used in thrusters and other devices in $\vec{E}\times \vec{B}$ fields. \par
In the present work a Penning type plasma source with a two cathodes, cube shaped anode and an electromagnet is used to study anode sheath . In this type of plasma, the electrons are confined electro-statically in axial direction while they are magnetically confined in radial direction. Hence forming a modified hollow cathode - Penning type of discharge. The electron sheath formation near the discharge electrode (i.e anode) itself and its evolution due to the magnetic field variation is studied. In the present set of experiments, no secondary electrode is used, which eliminates the need for an extra biasing power supply. The global response of the background plasma to the electron sheath formation is studied by using single Langmuir probe and Optical Emission Spectroscopy (OES). Experimental results on dynamics of plasma parameters, light emission and oscillations  are also described. Important results including the changes in confinement mode from ion sheath to electron sheath at critical magnetic field is presented in this work. Essential physics in this discharge revolves around the electron sheath and its evolution in to anode spot. The rest of the paper is organised as follows: the experimental setup  is described in Sec.~\ref{sec:setup}. The results for the evolution of anode glow, discharge current variation and changes in plasma properties  with magnetic field is discussed  in Sec.~\ref{sec:results}. At last a concluding remark is provided in sec.~\ref{sec:conclu}. 
\begin{figure*}
\includegraphics[width=.8\linewidth]{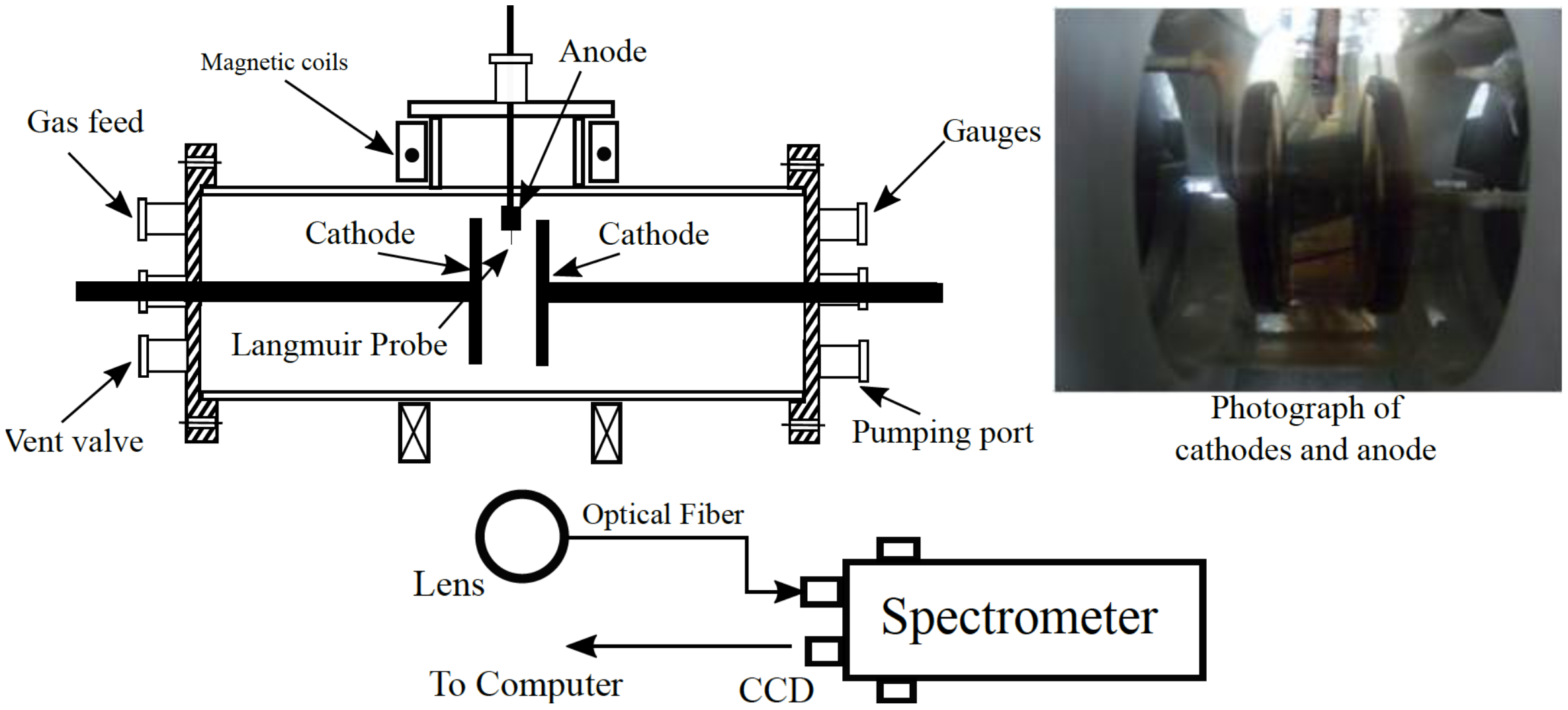}
\caption{Schematic of the experimental set up.}
\label{fig:setup}
\end{figure*}
\section{Experimental Setup}
\label{sec:setup}
The schematic of experimental set up is shown in fig. \ref{fig:setup}. It consists of discharge chamber of cylindrical glass tube with 200 mm diameter and 300 mm length. The stainless steel end flanges are provided with feed through for cathodes and ports for pumping and pressure gauge mounting. Two stainless steel circular discs of 100 mm diameter are used as cathodes. The stainless steel cathodes are covered with Teflon on the back side. Inter-cathode distance of 40 mm is kept fixed for all experiment. The stainless steel anode in the form of cube $(8\times 8\times 8\ mm^3)$ is used.Due to this shape of the anode, there is clear distinction of the electron flux reaching to the anode in the direction perpendicular to the magnetic field and parallel to the magnetic field. This grounded anode is placed symmetrically with respect to the two plane parallel cathodes. The anode is mounted through a port on the upper side of the chamber.The uniform axial magnetic field is produced by a pair of Helmholtz coils placed symmetrically relative to the centre of the vacuum chamber. The average distance between the two coils is 100 mm, equal to the radius of the coil. The magnetic field could be varied from 0 to 125 Gauss by varying the coil current. The chamber is evacuated by using rotary pump and diffusion pump to a base pressure of $10^{-5}\ mbar$. Then the working argon pressure is varied within the range of 0.03 to 0.125 mbar by feeding argon gas using a gas dosing valve. The power is provided by a 600 V and 1.5 A direct current power supply, which works in constant voltage mode. The two cathodes are connected to direct current power supply at the same potential while anode is grounded.  The discharge voltage is varied in the range of $300-600\ V$. The discharge voltage is measured using digital voltmeter of the power supply while current is measured using a series resistor in the circuit. \par 
A cylindrical Langmuir probe made up of tungsten with a tip length of 4 mm and 0.1 mm in diameter is used to measure the plasma properties near the anode.  The probe support is made up of ceramic tube which is fitted in a stainless steel rod. The grounded anode is used as a reference electrode for Langmuir probe. The Langmuir probe is always kept perpendicular to the magnetic field lines to minimise the effect of magnetic field on charge particle collection. Apart from probe measurement the images of the discharge near the anode were recorded by digital camera.  A high resolution spectrograph is used for spectroscopic diagnostics. The experimental arrangement is shown in fig. \ref{fig:setup}.  In the present set of experiments, optical emissions are collected at side-on locations (perpendicular to discharge axis ) using a light collection system. The optical probe has Plano convex lens (diameter: 50 mm and  focal length of 150 mm ) assembled in a small tube with a SMA (Sub Miniature version A) termination for connecting a 0.6 mm silica fiber (numerical aperture :0.22). In the experiment reported here the light collection system is placed such that it collects the radiations from the bulk plasma only. The spectra were recorded with a 0.5 meter imaging spectrograph (Acton research corp., USA) having a 1200 l/mm grating and equipped with a CCD (Charge Coupled Device) detector( Princeton instruments, USA). All the recordings were obtained by setting entrance slit at 50 micron for which the achievable wavelength resolution is $\sim\ 0.05$ nanometer. A high speed silicon photodiode (DET10A, Thorlabs) is  used for  spatially integrated light measurements.
\section{Results and Discussions}
\label{sec:results}
\subsection{Evolution of anode glow with magnetic field}
\begin{figure}[h]
\includegraphics[width=\linewidth]{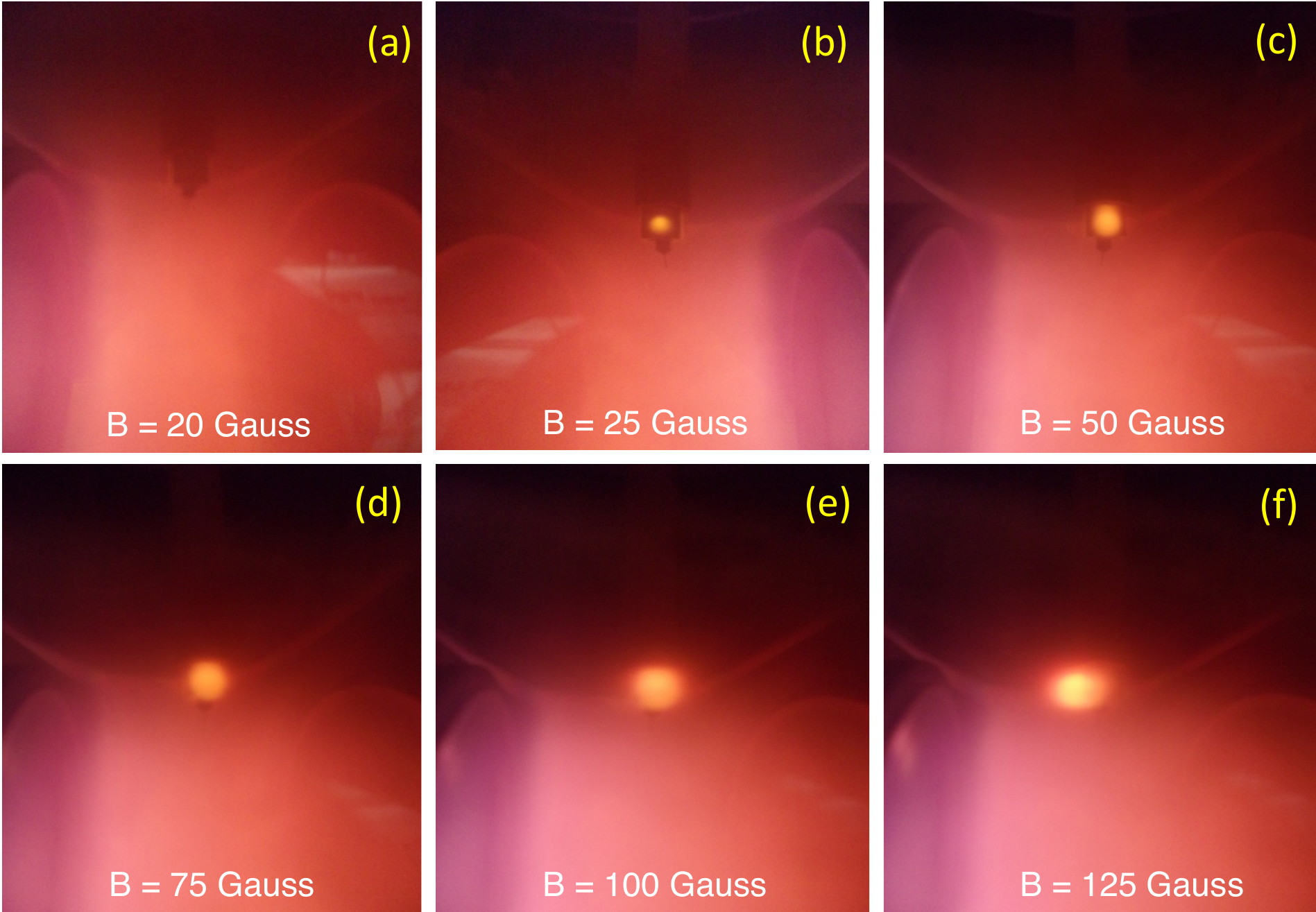}
\caption{Still images of the anode glow for different magnetic fields at discharge voltage of $500$~V and background pressure of $8\times10^{-2}$ mbar.}
\label{fig:Anode-pics}
\end{figure}

 The images of the anode glow at different values of magnetic field at a discharge voltage of $500\ V$ and background pressure of $8\times 10^{-2}\ mbar$ are shown in fig.\ref{fig:Anode-pics}. In the case of low magnetic field(i.e 10-15 Gauss), the anode does not show any glowing region. At this situation the anode collects required electrons from the plasma while retarding the rest, a typical ion sheath case. The condition is set by the balance between the electron flux at the anode and the current imposed on it by the external circuit. 
 When the  magnetic field is increased  to  20 Gauss, a faint glow near the anode starts appearing.  Further increase in magnetic field upto $25$~Gauss, results in  a clear formation of anode spot.Increasing magnetic field further results in larger anode spot (fireball) that covers the surface of the anode and extends out in the surrounding plasma . At magnetic field of 20 Gauss, the gyro radius of electrons is $\sim 1 mm$ (at 1 eV temperature) and gyro radius of ion is around 50 mm considering ions at room temperature. Though ions are not magnetized, electrons hold ions due to quasineutrality. 
As discussed in earlier work the anode spot or fireball formed is the result of the collisional ionization in the electron sheath  \cite{baalrud2009equilibrium, conde2006transition, chauhan2016droplet} ; hence, the electron sheath must form prior to the fireball. In such case, the electrons are accelerated in the electron sheath. If the potential is high enough it will ionize the background gas, the electrons so produced are quickly absorbed by the anode leaving behind the positive ion charge. Conde \textit{et.al.}\cite{conde2006transition} have shown that if the collisions are frequent it will result in neutralization of negative space charge hence, forming the glowing fireball region. Such fireball is usually separated from the bulk plasma by a double layer having potential equal to the ionizing potential of the gas\cite{gurlui2005dlpotential, baalrud2009equilibrium, song1991anode, chauhan2016droplet}. These visual observations of the anode sheath are justified in the following sections with discharge current, plasma potential and plasma density measurements of the bulk plasma.  
\subsection{Discharge Current variation with magnetic field}
\begin{figure}[h]
\includegraphics[width=1\linewidth]{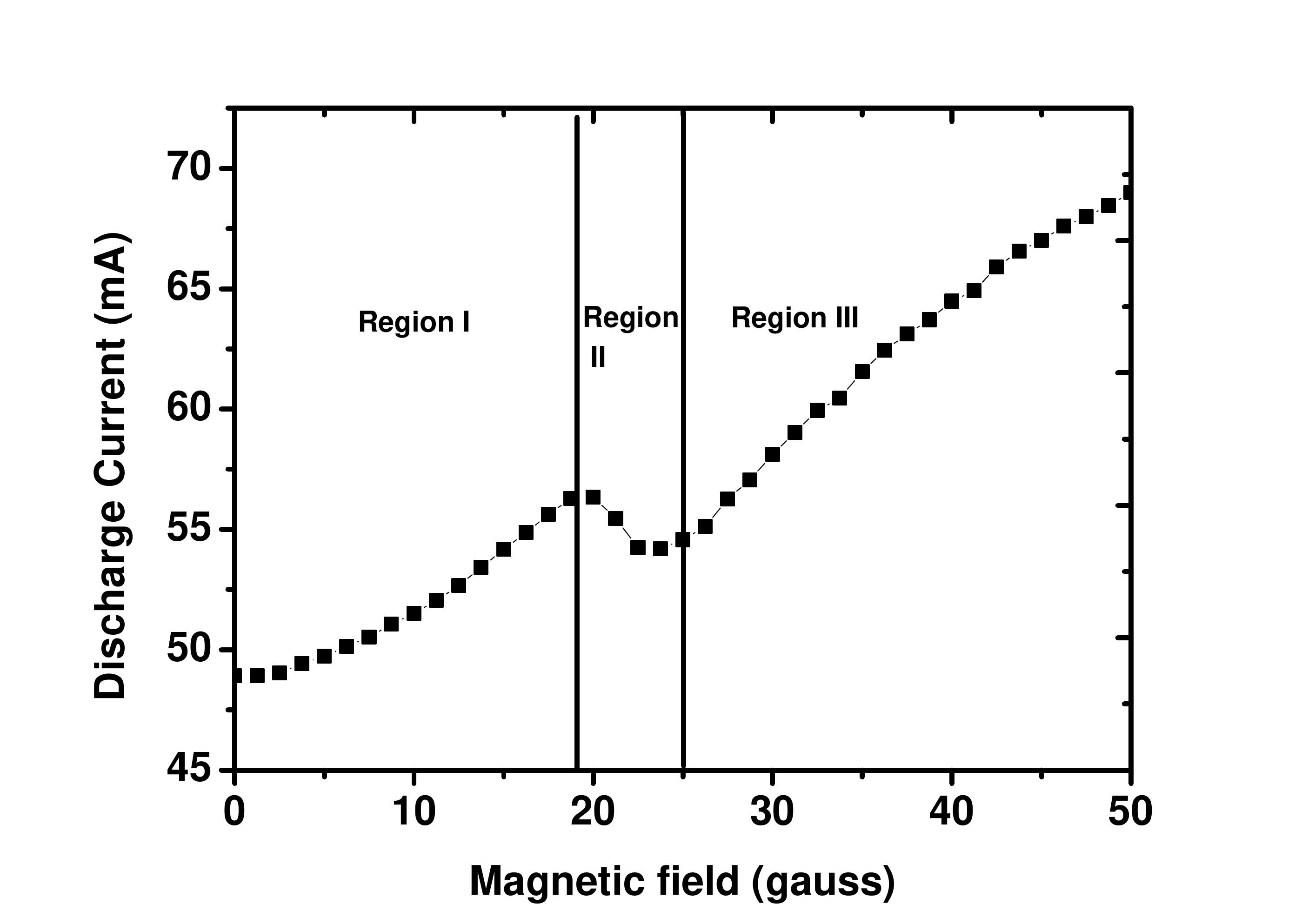}
\caption{The typical behavior of the discharge current with increasing magnetic field.}
\label{fig:typical}
\end{figure}

\begin{figure}[h]
\includegraphics[width=1\linewidth]{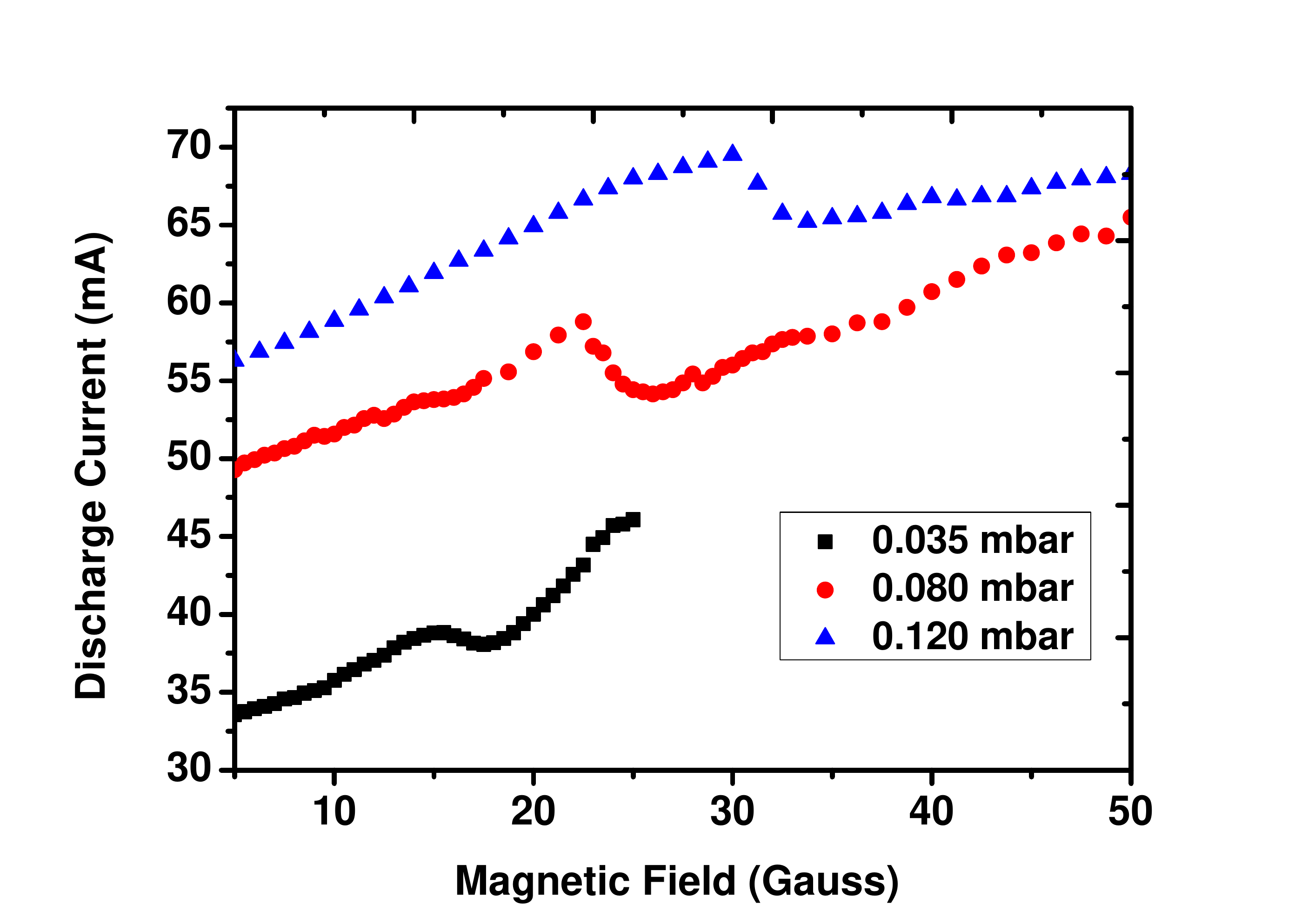}
\caption{Discharge current with increasing magnetic field for different background pressures at constant discharge voltage of 500 V.}
\label{fig:IvsP}
\end{figure}

\begin{figure}[h]
\includegraphics[width=1\linewidth]{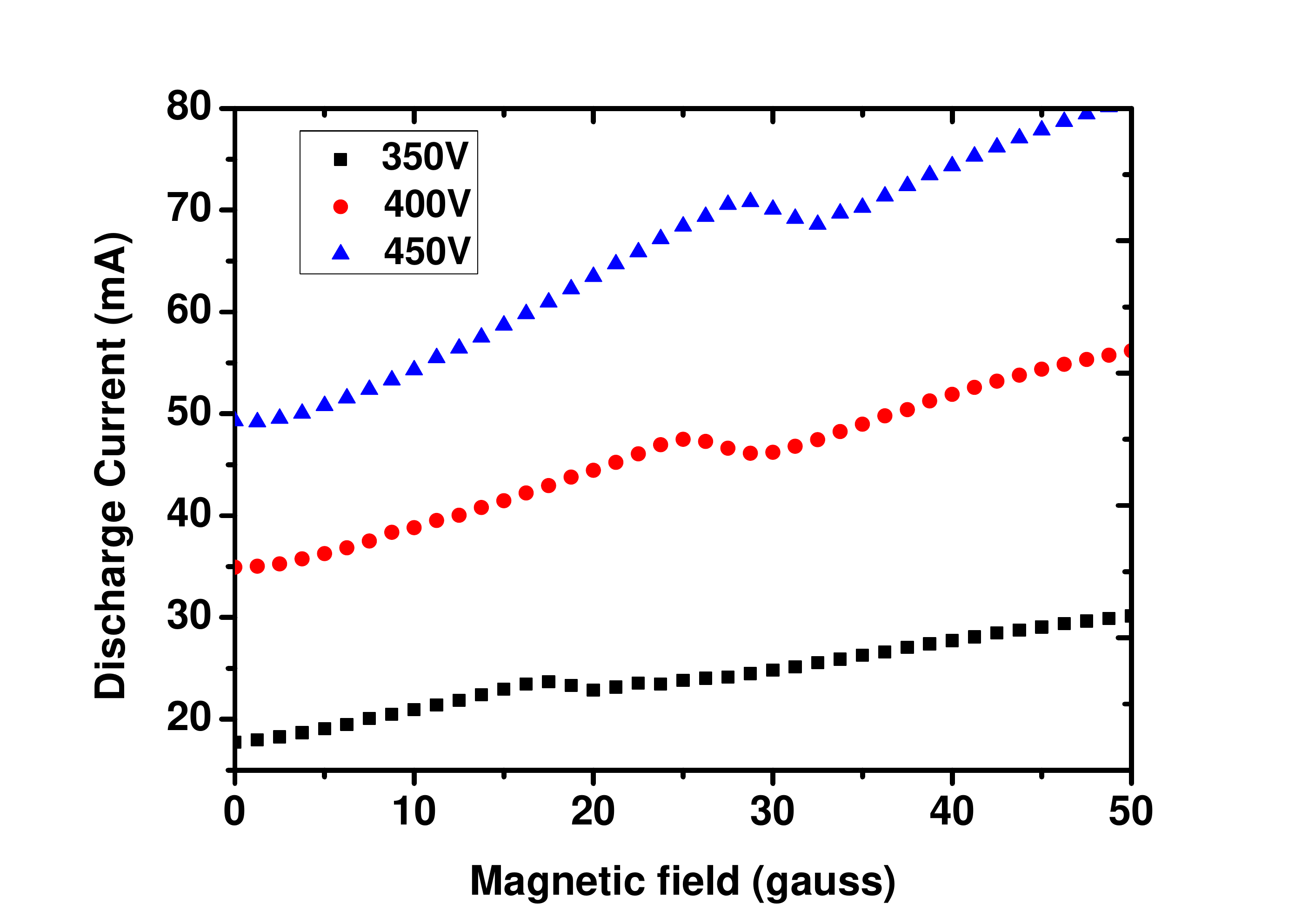}
\caption{Discharge current with increasing magnetic field for different discharge voltages at constant pressure of $10\times 10^{-2}$ mbar.}
\label{fig:IvsVd}
\end{figure}

The discharge current obtained at different discharge conditions is a very good indication of plasma source behaviour. Variation of the discharge current ($I_d$) with the magnetic field at pressure of $7\times 10^{-2}$ mbar and discharge voltage of $500\ V$ is shown in fig. \ref{fig:typical}. Initially at low magnetic field, the discharge current ($I_d$) shows an increasing linear trend  (i.e. region I of fig \ref{fig:typical}) as magnetic field is increased. This increase in $I_d$ is a result of better confinement and increased collisions due to cyclotron motion of electrons which increases overall electron and ion density inside the plasma. This nearly linear increase in $I_d$  however stops at about 20 Gauss and the trend reverses. The discharge current decreases on further increase of magnetic field up to 25 Gauss (region II, fig \ref{fig:typical}) before again showing positive trend (region III of fig \ref{fig:typical}). Region III of fig \ref{fig:typical} shows localized glow known as anode spot (see fig \ref{fig:Anode-pics}). The above transition between different regions was also studied at different discharge conditions. Fig. \ref{fig:IvsP} shows the variation of discharge current for different pressure. The critical magnetic field required to induce transition from region I to region II and III increases with increasing pressure. This suggests that the  ratio of electron neutral collision mean free path ($\lambda_{en}$) to the electron larmor radius ($r_{L}$)has a role to play. Similarly, fig. \ref{fig:IvsVd} shows the variation of the discharge current at different discharge voltage for given pressure. In this case as well the critical magnetic field for transition rises with applied voltage.The increase in pressure as well as increase in voltage results in higher electron-neutral collision frequency as well as discharge current. The electrons will be magnetised sufficiently when electron cyclotron frequency is greater than the collision frequency. Hence it can be qualitatively  said that the higher magnetic field is required for the transition from region I to region II.
\subsection{Plasma Potential and Floating Potential}
\begin{figure}[h]
\includegraphics[width=1\linewidth]{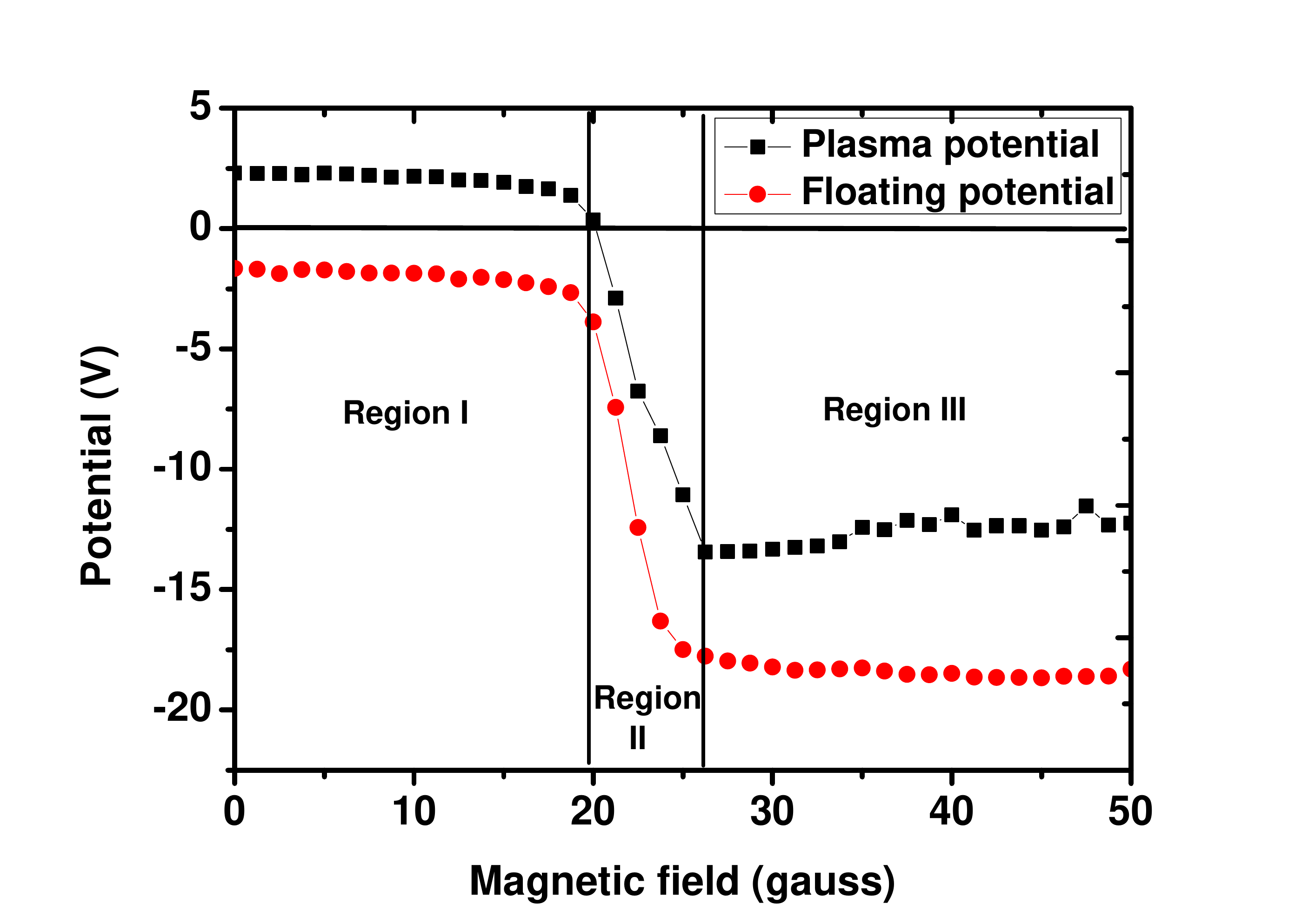}
\caption{Variation of plasma potential and floating potential in the bulk plasma with increasing magnetic field at constant voltage of 500 V and pressure $7\times 10^{-2}$ mbar.}
\label{fig:Vp_vf_vsB}
\end{figure}

In order to measure the plasma potential and floating potential, a Langmuir probe is placed near the anode at 10 mm distance from the anode surface. The probe bias voltage at which the second derivative of probe current becomes zero is considered as plasma potential whereas, the floating potential is a potential at which the net flux of electrons and ions to the probe becomes equal and hence net current collected by the probe becomes zero. The plasma and floating potentials are measured for different magnetic field at constant discharge voltage and plotted in fig. \ref{fig:Vp_vf_vsB}. It is observed that in region I i.e. upto the magnetic field of 20 Gauss, the plasma potential is few volts positive ($\sim\ 2\ V$) above the anode potential. The plasma  potential and floating potential remain almost constant upto the  field of 20 Gauss. Beyond 20 Gauss of magnetic field the plasma potential systematically decreases (region II) and becomes constant again (region III) for larger values of the magnetic field. These observations can be explained as follows.\par
Plasma potential of bulk plasma (measured with respect to the anode) indicates the type of sheath formed at the anode surface. In case of a typical ion sheath, the anode (or any other electrode) is negative with respect to the plasma potential. This is the case in the region I of fig. \ref{fig:Vp_vf_vsB}. Formation of ion sheath is a result of the balance between discharge current and electron flux at the anode. In simplest case it can be argued that the ion sheath forms when the thermal electron flux ($J_e$) at anode is larger than the discharge current ($I_d$), i.e. $AJ_e>I_d$, where $A$ is the area of the anode. Similarly, when $AJ_e<I_d$ is satisfied the electron sheath appears at the anode. In such cases the plasma potential becomes negative with respect to the anode (or any other electrode). This situation arises in region II and III in fig. \ref{fig:Vp_vf_vsB}. Very interestingly the potential difference between the anode and the plasma is $14\ V$ which is close to the ionization potential of the argon. This has been verified by using the Helium gas also, for which the potential difference is measured to be $\sim\ 24\ V$. It has been reported numerous times that in case of the formation of fireball at the anode the potential difference between the anode and plasma is always close to the ionization potential of the gas used in the discharge\cite{gurlui2005dlpotential,baalrud2009equilibrium,song1991anode,chauhan2016droplet}. The visual observation of the anode beyond $25$~Gauss of applied magnetic field confirms the presence of localized bright anode spot or fireball. Anode fireball forms as the result of the ionization due to electrons accelerated in the electron sheath, hence the prior condition is the presence of the electron sheath.
Putting everything in order, increasing the magnetic field increases the discharge current up to a certain value, then the discharge current decreases for up to few Gauss at which the potential profile suggest the evolution of the electron sheath from the earlier ion sheath. Finally, the fireball appear at the anode which enhances the electron collection area of the anode and the discharge current again starts increasing due to combined effect of increased ionization and increased collection area.\par
To understand the role of magnetic field in the transition from ion sheath to electron sheath, we can evoke the model used by Baalrud \textit{et.al.}\cite{baalrud2009equilibrium} The criterion for formation of particular type of sheath as discussed above can be written in more useful way as the  ratio of the electron to ion collecting surface areas. This is because the particle flux to any surface is a function of the local density and temperature of that particle. Building upon this fact, Baalrud \textit{et.al.} have concluded that the ion sheath will form when the ratio $A_e / A_i > 1.7\sqrt{2.3m_e/m_i}$, where $A_e$ and $A_i$ are electron and ion collection area, respectively and $m_e$ and $m_i$ are electron and ion mass. For electron sheath formation, the condition turns out to be $A_e / A_i < \sqrt{2.3m_e/m_i}$. \par
In absence of the magnetic field the whole cube of the anode collects the current. And hence, the area of full anode has to be considered. In this case the area ratio ($A_e / A_i $) is $17\times 10^{-3}$ which is greater than that required for ion sheath formation and hence an ion sheath would form. Now, applying the magnetic field results in better confinement of the electrons and hence the wall loss reduces hence the discharge current increases initially. However once the electrons are sufficiently magnetized they follow the field lines and misses the anode except for the two sides where the magnetic field penetrates the anode. This effectively reduces the area of the anode to $128\ mm^2$ and the area ratio becomes $8\times 10^{-3}$ which satisfy the criterion for electron sheath formation. Hence the experimental observation can be explained by the reduction of effective anode collection area due to increased magnetic field.
\subsection{Electron Temperature and Plasma Density}
\begin{figure}[h]
\includegraphics[width=1\linewidth]{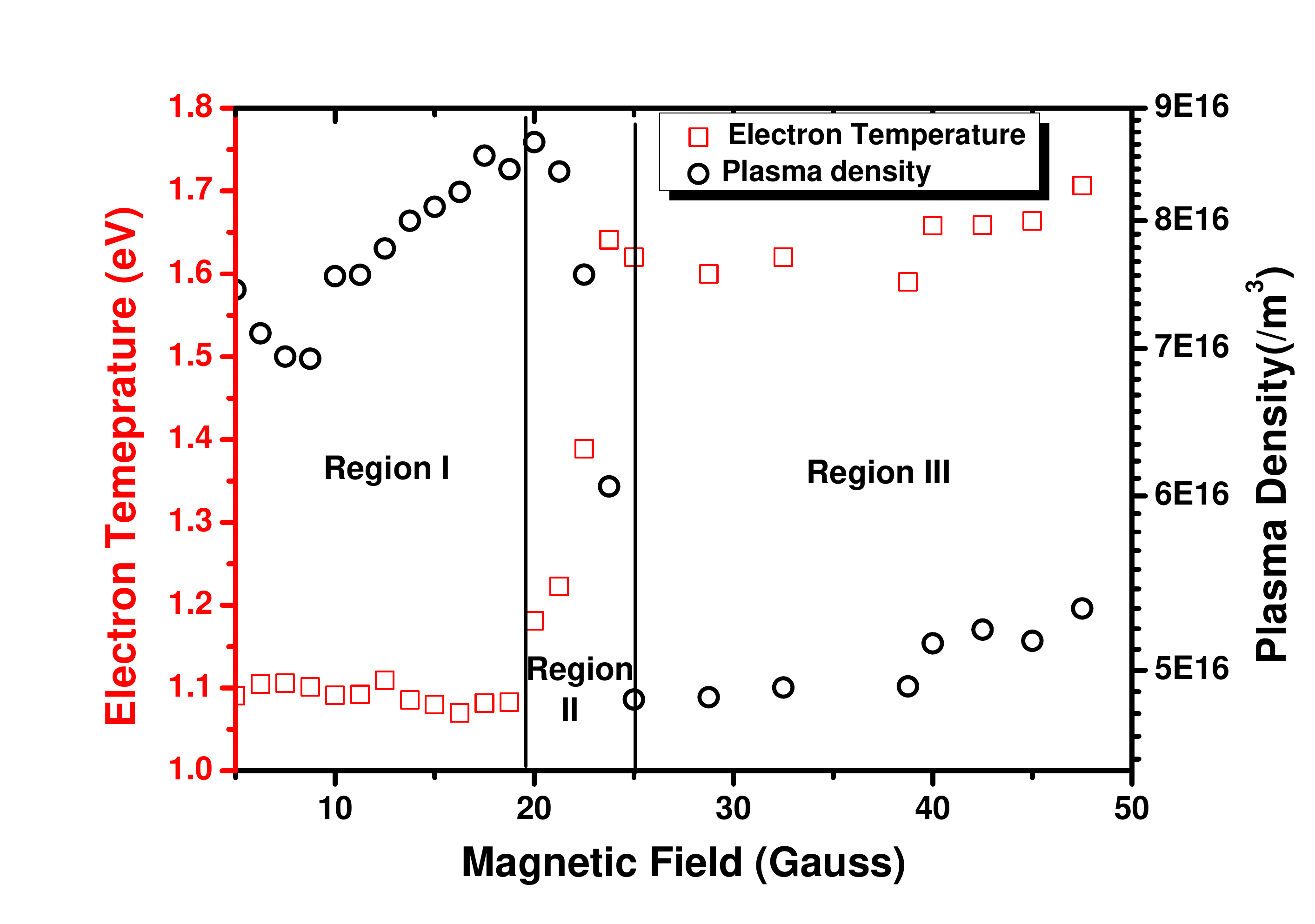}
\caption{Variation of bulk plasma density and electron temperature with increasing magnetic field.}
\label{fig:Te_Ne_vsB}
\end{figure}

The changes in the background plasma density and electron temperature are measured using a single Langmuir probe. Fig. \ref{fig:Te_Ne_vsB} shows plasma density and electron temperature at different applied magnetic field for discharge voltage of $500\ V $ and background pressure of $7\times 10^{-2}\ mbar $. It is observed that electron temperature increases in region II and III. The increase in electron temperature due to the presence of the electron sheath has been reported earlier \cite{baalrud-2007-global}. It can be explained as follows. In absence of electron sheath and low magnetic field in region I, the plasma is bounded by ion sheath on every surface. By nature, ion sheath have negative potential with respect to the plasma. This structure hence reflects electrons, only electrons having sufficient energy to overcome the barrier are lost to the walls. Hence, the high energy electrons are selectively removed from the system, this results in over population of low energy electrons in the plasma leading to lower electron temperature. In region II and III; however, the plasma - anode interface is electron sheath and hence all of the electrons can be lost to it without regard to its energy. Additionally in the radial direction the electron loss is greatly reduced due to the magnetic field. This mitigates the effects that existed in region I, leading to increased temperature due to better confinement of electrons. 
Fig. \ref{fig:Te_Ne_vsB} shows reduction in plasma density when electron sheath forms near anode. As soon as electron sheath formation takes place, the electrons from the global plasma  gets drained to the anode. There is a severe electron loss to the anode. Hence the bulk plasma density decreases. Similar type of plasma density reduction in response to the electron sheath formation on the  positively biased plate has been observed\cite{Bailung2017} earlier. The steady increase in the plasma density in region III can be attributed to increased ionization in anode spot and better confinement by magnetic field. The electron Larmor radius for $1\ eV$ electron and $20$ and $40$~Gauss of magnetic field is $1.2$ and $0.6\ mm$ respectively; which is much smaller than typical electron - argon mean free path of $10\ mm$ at $10^{-2}$ mbar. Though distance between cathodes is 40 mm the anode cathode distance is $20\ mm$ and taking into account the size of anode $8\ mm$ and the thickness of cathode sheath (visually observed) $\sim 5\ mm$, the distance between cathode sheath edge and anode is $\sim 10\ mm$. Hence, the confinement effect can be well observed at applied magnetic field of $20-40$~Gauss.\par
The behaviour of probes in magnetic field could sometimes be erroneous, proximity of the probe to the anode may also affect the result to some extent. Hence the reduction in the plasma density and increase in plasma temperature can be confirmed with an independent non-intrusive diagnostics. Hence optical emission spectroscopy (OES) of the discharge is carried out to complement the results obtained by the probe. The results of OES are described in the following  section.
\subsection{Optical Emission Spectroscopy diagnostic of global plasma}
\begin{figure}[t]
\includegraphics[trim=0 7cm 0 7cm,clip,width=0.9\linewidth]{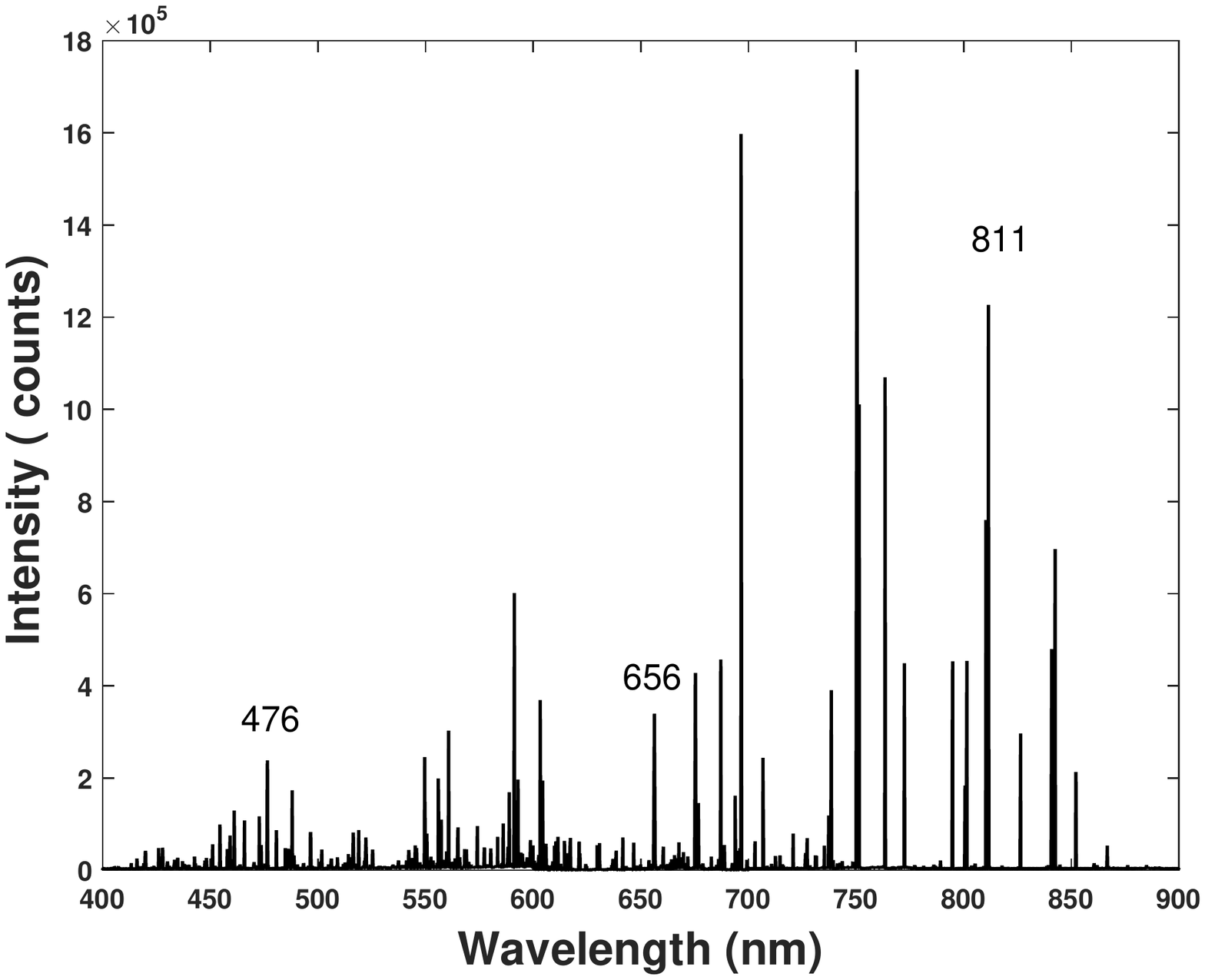}
\caption{Typical Argon spectra showing  Ar*II, H* and Ar*II emission lines.}
\label{fig:spectra}
\end{figure}
\begin{figure}[t]
\includegraphics[width=1\linewidth]{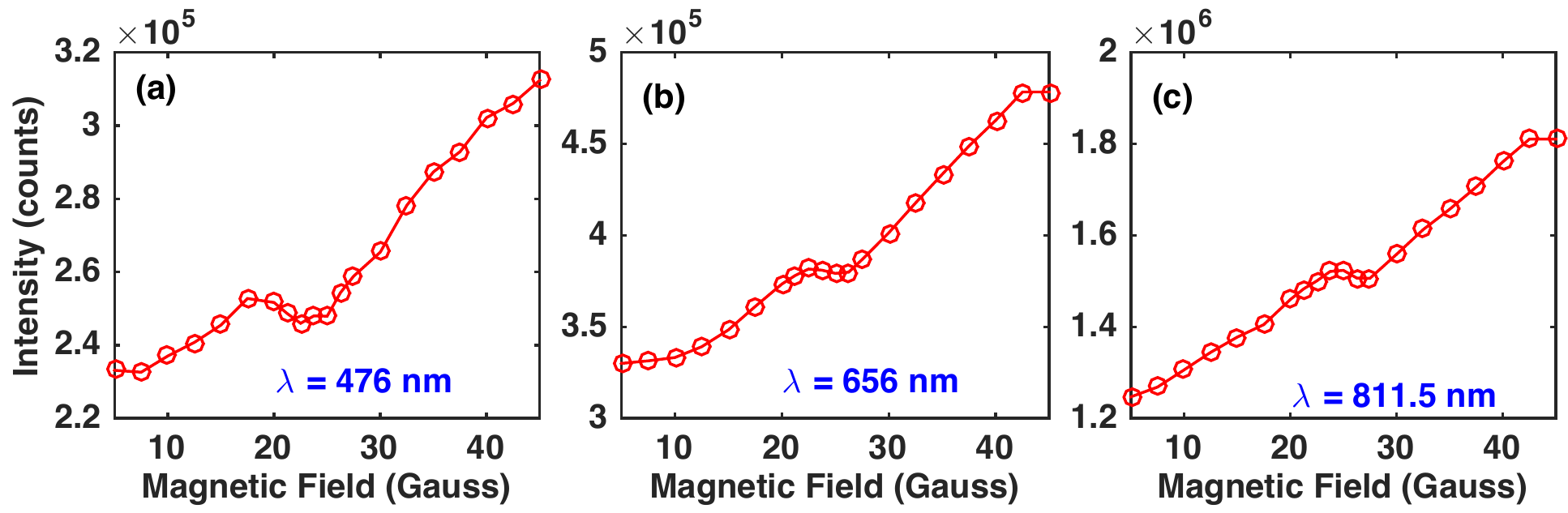}
\caption{Line Intensity variation of (a) $476\ nm$ Argon II line (b) $656\ nm$ Hydrogen I line and (c) $811.5\ nm$ Argon I line with increasing magnetic field.}
\label{fig:lines}
\end{figure}
The emissions are observed with the help of a high resolution spectrograph and the argon spectra (400 nm to 900nm) are obtained at applied voltage of $500$~V and pressure of $6-9\times 10^{-2}$~mbar for a range of magnetic field from 0 to 50 Gauss. The example spectra is shown in the Fig. \ref{fig:spectra}. Intense neutral argon lines (Ar I) as well as ionic argon lines ( Ar II) are observed, apart from this spectral lines corresponding to excited neutral hydrogen are also observed. As the pressure is very low, assuming that the argon atoms are excited by electron impact excitation and decayed radiatively by spontaneous decay, it can be shown that the observed intensity and hence the density of excited neutrals depend on plasma density and very weakly  influenced by plasma temperature via reaction rate\citep{Griem}. In such case, observed intensity of the spectral lines should decrease in region II of Fig. \ref{fig:Vp_vf_vsB} and Fig. \ref{fig:Te_Ne_vsB}. For this purpose, the intensity of spectral lines of excited neutral argon lines( Ar I: 811 nm, transition $3p^5({}^2P_{3/2})4s - 3p^5({}^2P_{3/2})4p$ ), ionic argon line (Ar II: 476 nm, transition $3s^23p^4(3_P)4s - 3s^23p^4(3_P)4p$) and also the Balmer alpha line(H I: 656.3 nm, transition $2p - 3d$) are chosen and the intensity variation with applied magnetic field is plotted in Fig. \ref{fig:lines}. \par
It can be observed that even if the spectra are from three different species i.e ( Ar*I, Ar*II and H*), the  intensity of these spectral lines increase  initially  in region I and  decrease  in  region  II  and  again start to  rise  in region III confirming the observations shown in earlier sections. The electron temperature and density can be obtained by coronal modelling of the observed intensities and can be compared with the results obtained by Langmuir probe. The work is under progress and  will be published in a separate manuscript.
\subsection{Spatio-temporal measurements}
\begin{figure}[h]
 \begin{center}
\includegraphics[width=1.0\linewidth]{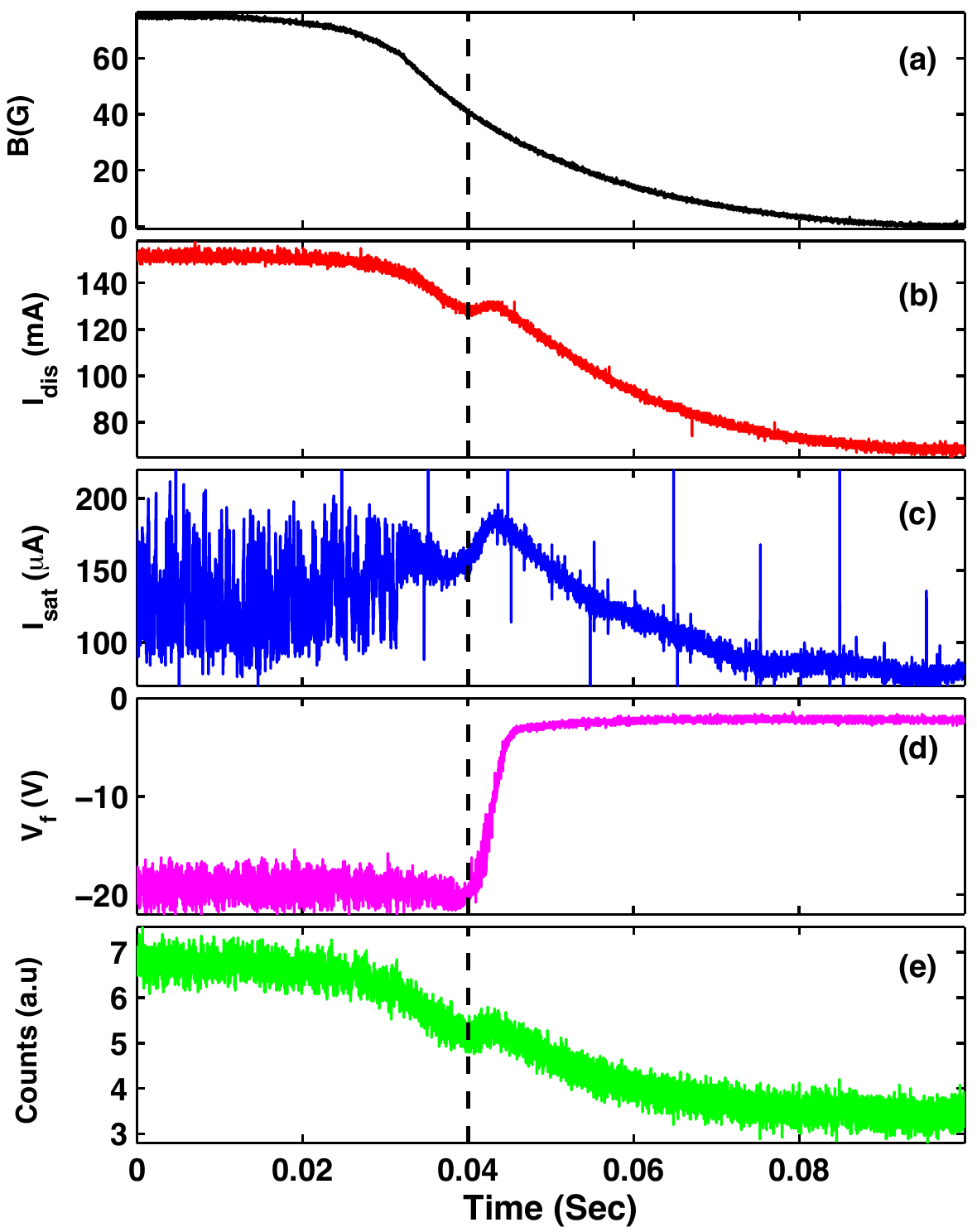}
\caption{The real time signal of (a) Magnetic field (b) Discharge current (c)Ion saturation current (d)Floating potential and (e) Intensity of light at  600 V and pressure $9\times 10^{-2}$ mbar.}
\label{fig:Fig10}
\end{center}
\end{figure}

As discussed in previous sections, the  various plasma  parameters  like  plasma potential, plasma density, electron temperature as well as intensity of emission lines are influenced during the anode sheath transition from ion sheath to electron sheath. However, the temporal variation of these parameters, as shown in Fig.~\ref{fig:Fig10}, with magnetic field during this transition give more insight of the phenomenon taking place in the discharge. In order to capture the transition phenomenon, the fast variation of magnetic field from 75 Gauss to 0 Gauss in 100 ms (as shown in Fig.~\ref{fig:Fig10}(a)) is carried out by switching off the coil current abruptly. The discharge current (shown in Fig.~\ref{fig:Fig10}(b)), ion saturation current (shown in  Fig.~\ref{fig:Fig10}(c)), floating potential (shown in  Fig.~\ref{fig:Fig10}(d)) as well as integrated intensity of spectral lines (shown in  Fig.~\ref{fig:Fig10}(e)) are measured simultaneously during the magnetic field variation at discharge voltage of 600V and operating pressure of $9\times 10^{-2}$ mbar. The high speed silicon photodiode (DET 10A) is used to detect the  integrated light signal of the emitted argon spectral lines. The spectral response of this diode is such that its responsivity is highest for the required wavelength range of 600-850 nm  in which intense argon  spectral lines were recorded using spectrometer as shown in Fig.~\ref{fig:spectra}. The ion saturation current and floating potential are recorded by single Langmuir probe which essentially indicate the variation of plasma density and plasma potential, respectively.\par
 \begin{figure}[h]
 \begin{center}
\includegraphics[width=0.9\linewidth]{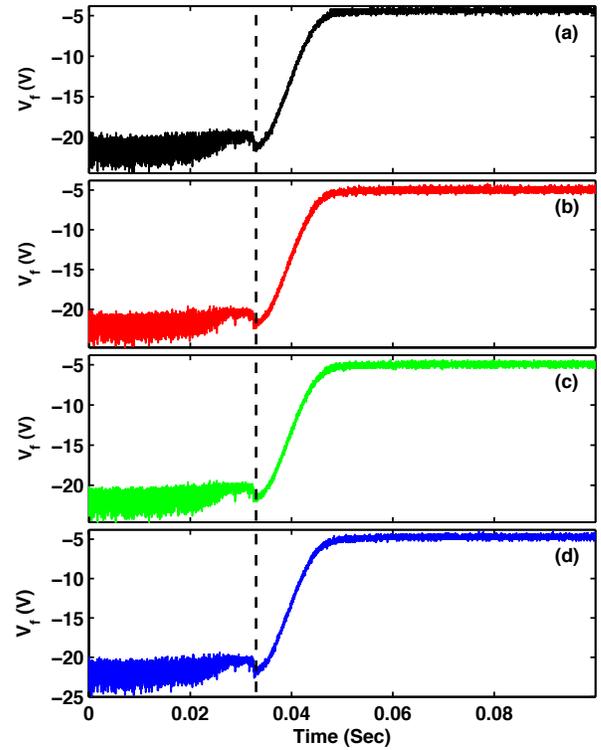}
\caption{The real time signal of floating langmuir probes at distance of (a) 10 mm (b) 15 mm (c) 20 mm and (d) 25 mm from the anode during the magnetic field variation}
\label{fig:Space variation}
\end{center}
\end{figure}
It is clear from the Fig.~\ref{fig:Fig10}, the discharge current, ion saturation current as well as intensity of emitted spectral lines shows similar behaviour during the decay of magnetic field from 75 Gauss to 0 Gauss. Even though the magnetic field is continuously decreasing, at a critical magnetic field of $\sim 40$ Gauss, all these three parameters shows characteristics dip (slightly increases before decreasing again)  when the transition from electron sheath to ion sheath takes place. At a given discharge parameters, the discharge current and the ion saturation current are the direct measure of plasma density. The population of excited species of argon neutrals and hence intensity of emitted argon lines depends on the plasma density\citep{Griem}.
 Hence, it can be concluded that with the decrease of magnetic field, the plasma density decreases and hence all these parameters like discharge current, ion saturation current and the line intensity follow the same trend of magnetic field except at the time when the transition takes place. It is also worth to mention that during the transition of electron sheath to the ion sheath, the spectral line intensity momentarily increases due to the increase of plasma density which essentially follows the same trend of ion saturation current and the discharge current. During the transition, the floating potential suddenly jumps to a higher value, otherwise it is locked to a fixed value in the electron sheath region. All these findings are in line with the steady state measurement of plasma parameters that are discussed in subsections of Sec.~\ref{sec:results}\par
In order to study the spatial variation of this response in the plasma, the floating potentials are measured simultaneously using four identical Langmuir probes which are placed at a distance of 5 mm from each other. These probes are mounted below the anode in a direction perpendicular to the magnetic field. Fig.~\ref{fig:Space variation} shows that all the four probes show the same behaviour of floating potential during the magnetic field variation. These measurements essentially confirm that the spatial variation in the response of the plasma parameters during the sheath transition is insignificant.
\subsection{Anode glow Oscillations}
\begin{figure}[h]
\includegraphics[width=\linewidth]{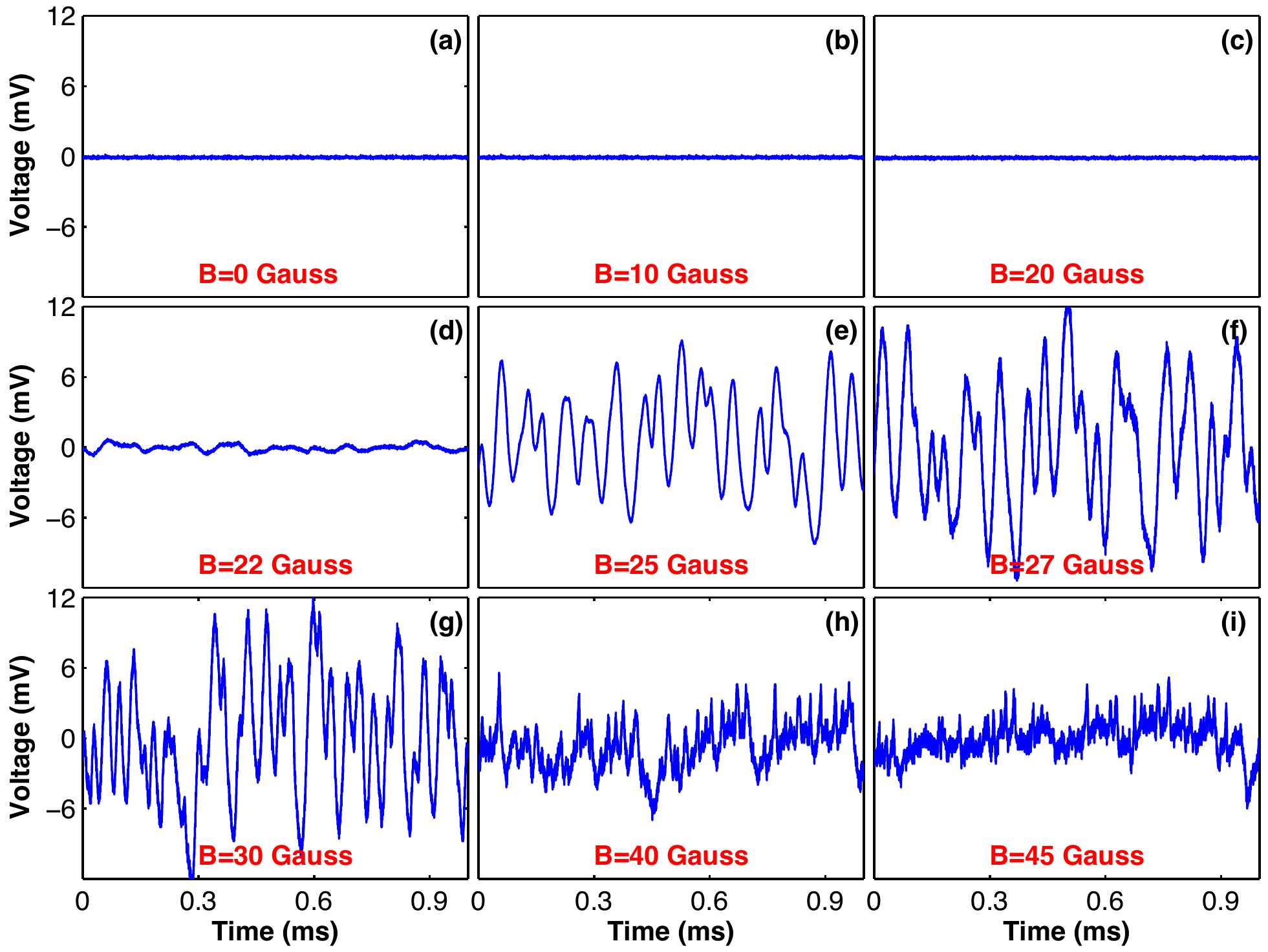}
\caption{Oscillations of floating potential observed at various magnetic fields.}
\label{fig:Osci}
\end{figure}

In order to study the stability of the discharge during the electron sheath formation, the temporal variation in floating potential is recorded. The fig. \ref{fig:Osci} shows the time series of floating potential signal for different magnetic field. 

\begin{figure}[h]
 \begin{center}
\includegraphics[width=1.0\linewidth]{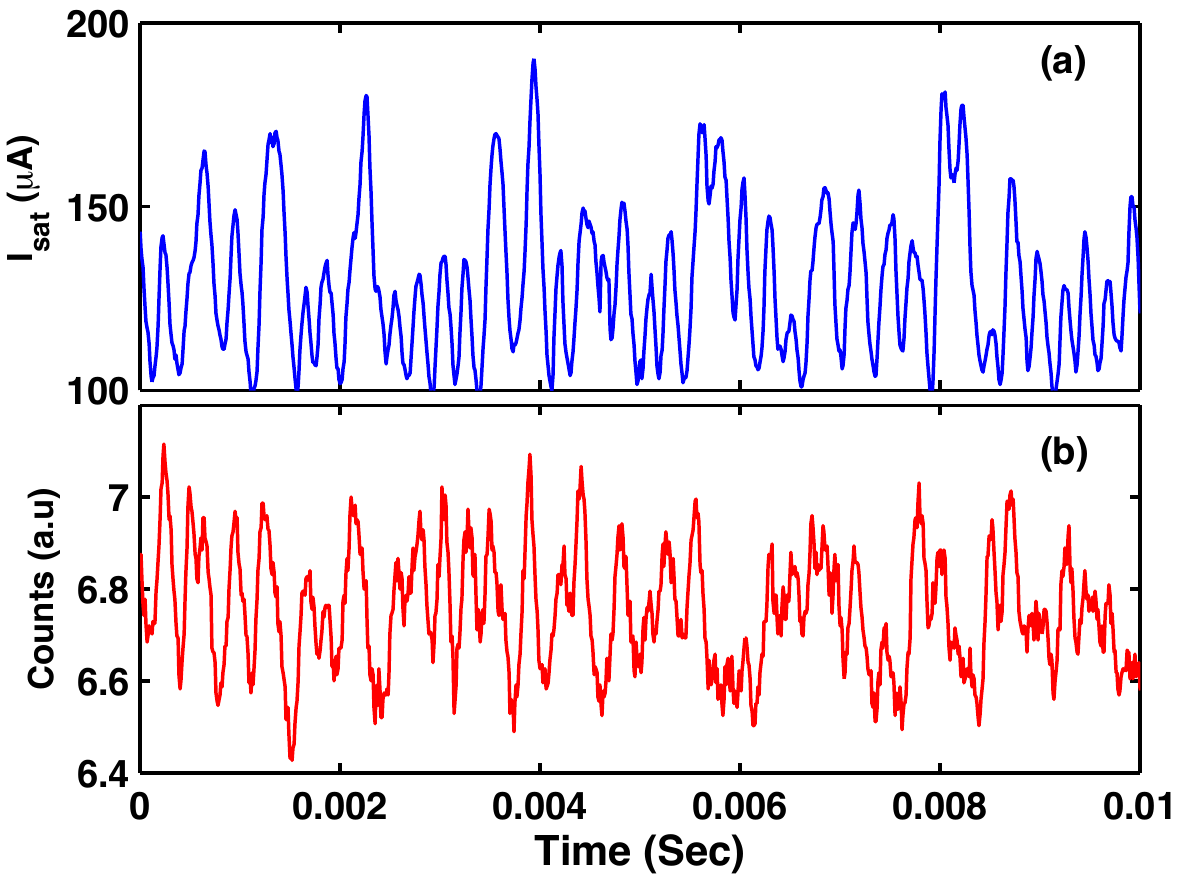}
\caption{The fluctuations in (a)Ion saturation current and (b) Light Intensity at magnetic field of 50 Gauss }
\label{fig:oscillations}
\end{center}
\end{figure}
It is clear from the figure that the discharge is stable upto 20 Gauss and the floating potential signal does not show any fluctuations. However, as soon as the electron sheath formation takes place, the oscillations in the global discharge parameters like floating potential and discharge current are observed. The oscillation frequency is around 5-20 kHz. The absence of oscillations in low or zero magnetic field and it appearance with the onset of the electron sheath and fireball suggest its origin to be in the electron sheath and ionization in it. These types of oscillations are well reported in literature in presence of fireball or anode spot\cite{song1991anode,karkari,chaubey2015sync} and is understood to be the result of repetitive formation and collapse of the double layer structure of the anode spot. The  similar oscillations are also observed in the light intensity of the fireball and ion saturation current when the electron sheath is converted into anode spot at higher magnetic field. Fig.~\ref{fig:oscillations} shows the oscillations in light intensity and ion saturation current at a given magnetic field and discharge parameters. It is to be noted that the  trend  of the oscillations are similar in both the signals. This suggests that the repetitive growth and decay of the fireball strongly depends upon the density replenishment of the background plasma.
\subsection{ Theoretical analysis for the electron flux near the anode}
\begin{figure}[h]
\includegraphics[width=\linewidth]{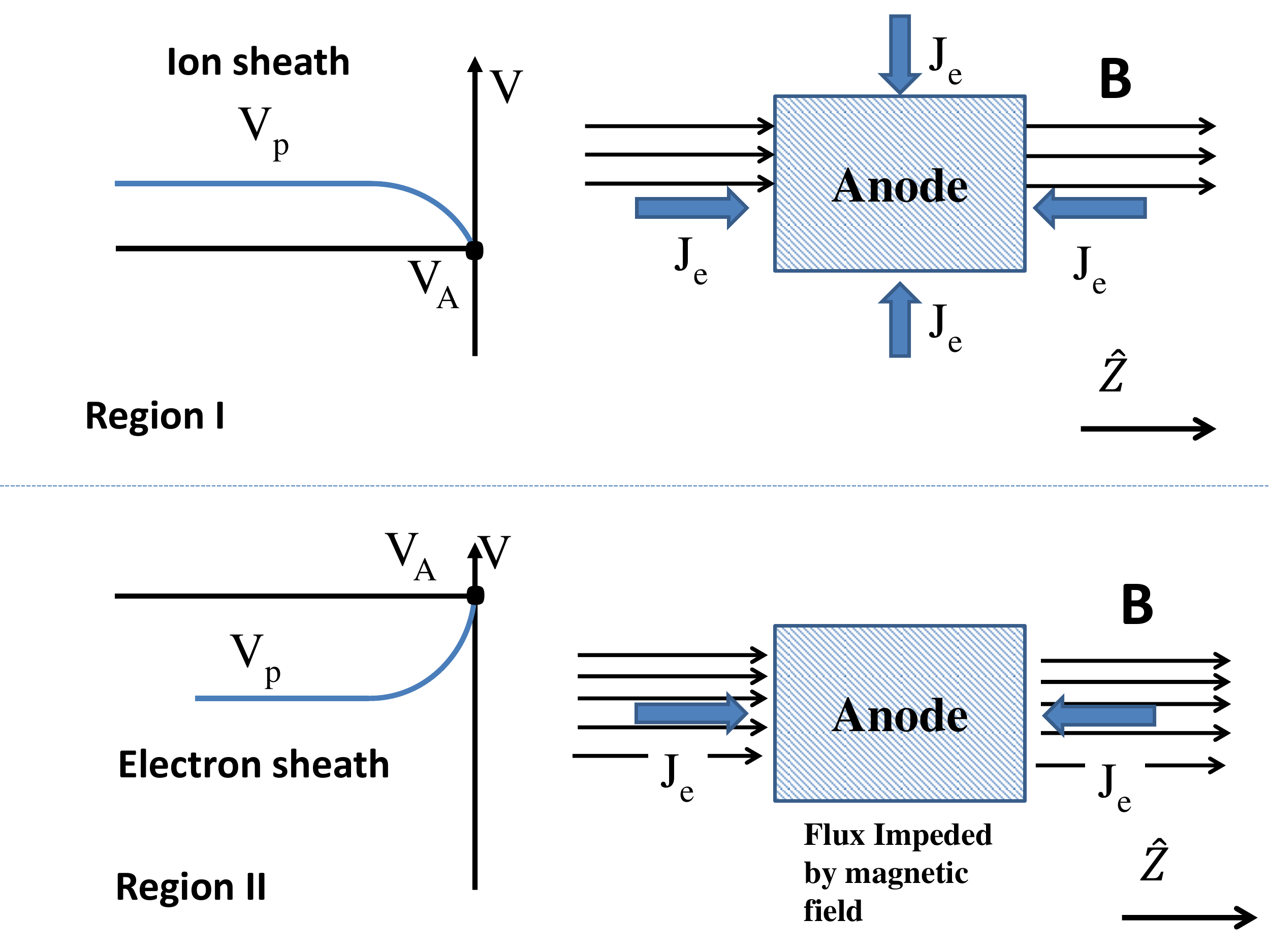}
\caption{Schematic representation of the perpendicular and parallel electron fluxes entering to the anode sheath.The region I represents low magnetic field where electron sheath is not formed while region II represents electron sheath formation in high magnetic field (i.e.greater than 20 Gauss)}
\label{fig:Sch}
\end{figure}
A schematic representation of the anode surface and particle flow in region I and region II is shown in Fig.\ref{fig:Sch}. The particle i.e electron  flow or flux is parallel to the magnetic field for two surfaces of the anode while it is perpendicular to the magnetic field for remaining surfaces. In a weakly ionised plasma,the diffusion coefficients for particle flow parallel and perpendicular to the magnetic field are expressed as.
\begin{eqnarray}
D_{\parallel}=\frac{KT_e}{m\nu_c} 
\end{eqnarray}
\begin{eqnarray}
D_{\perp}=\frac{D_{\parallel}}{1+\omega^2/\nu_c^2}
\end{eqnarray}
where $D_\perp$ and $D_\parallel$ are the perpendicular and parallel diffusion coefficients. $m$ is the mass of electron, $KT_e$ is electron temperature. The $\omega=eB/m$ is electron cyclotron frequency while $\nu_c$ is electron neutral collision frequency. The electron flux normal to the field lines is governed by the expression
\begin{eqnarray}
\frac{J}{e} = - D_{\perp} \frac{\partial n}{\partial z}
\end{eqnarray}
In the region I, the electron flux reaching to the anode from all the sides is not disturbed. Thus the thermal electron flux to the anode would carry a current much greater than discharge current. In fact considering the full anode area, the thermal electron current $\sim100 mA$ calculated from measured plasma density $\sim 10^{16} m^{-3}$ and electron temperature $\sim ~1 eV$ is higher than the discharge current $\sim 50 mA$. Therefore the formation of  electron retarding sheath is required to repel excessive electron flux to the anode so that global current continuity is maintained. However, in region II, the electron current reaching to the anode is impeded due to the transverse magnetic field. It is clear from equation (2) and (3)that the diffusion coefficient is going to decrease with the magnetic field and hence the electron current reaching to the anode. In our experimental parameter i.e magnetic field $20-50$~Gauss and pressure $ 10^{-2}$-$ 10^{-1}\ mbar$ , the ratio $\omega^2/\nu_c^2$ is in the range of 1 to 2. It is expected that the electron flux  reaching to the anode should reduce considerably in region II. In other words it can be said that when the electron flow is blocked to the some of the anode surface, the discharge current should close to the anode at the surface where the electron flow is parallel to the magnetic field. However a total thermal electron current towards this surface is smaller than the discharge current. The additional electron flux is attracted towards the anode by the electron attracting sheath that appears at the surface. The discharge current closes to the anode via this electron sheath and additional ionisation in it.
\section{Conclusion}
\label{sec:conclu}
Magnetic field is shown to induce the formation of the anode spot or fireball on the surface of the anode. This happens primarily due to restricted anode area. The discharge current and emission spectra shows similar behaviour validating the arguments. The anode spot is observed to induce global oscillations in plasma parameters as reported previously by others. Measured plasma potential clearly shows that upon increasing magnetic field the anode remains below the plasma potential initially suggesting an ion sheath. However, further increasing the magnetic field, it drops below the plasma potential by about the same value as the ionization potential of the gas, suggesting anode spot or fireball at the anode surface. The transition can be understood as the result of the restricted electron flow to the anode, which rises the anode potential leading to the formation of electron sheath and eventually to anode spot or fireball. Observed trend of plasma density and electron temperature in the transition region is the result of the formation of the anode spot, which changes local plasma parameters. The real time measurements of the discharge  parameters like discharge current, plasma density, intensity of emitted argon lines also shows a dip during the transition of ion sheath to electron sheath, which suggests that the transition of anode sheath produces a density and potential variations into the background plasma.

\begin{thebibliography}{26}%
\makeatletter
\providecommand \@ifxundefined [1]{%
 \@ifx{#1\undefined}
}%
\providecommand \@ifnum [1]{%
 \ifnum #1\expandafter \@firstoftwo
 \else \expandafter \@secondoftwo
 \fi
}%
\providecommand \@ifx [1]{%
 \ifx #1\expandafter \@firstoftwo
 \else \expandafter \@secondoftwo
 \fi
}%
\providecommand \natexlab [1]{#1}%
\providecommand \enquote  [1]{``#1''}%
\providecommand \bibnamefont  [1]{#1}%
\providecommand \bibfnamefont [1]{#1}%
\providecommand \citenamefont [1]{#1}%
\providecommand \href@noop [0]{\@secondoftwo}%
\providecommand \href [0]{\begingroup \@sanitize@url \@href}%
\providecommand \@href[1]{\@@startlink{#1}\@@href}%
\providecommand \@@href[1]{\endgroup#1\@@endlink}%
\providecommand \@sanitize@url [0]{\catcode `\\12\catcode `\$12\catcode
  `\&12\catcode `\#12\catcode `\^12\catcode `\_12\catcode `\%12\relax}%
\providecommand \@@startlink[1]{}%
\providecommand \@@endlink[0]{}%
\providecommand \url  [0]{\begingroup\@sanitize@url \@url }%
\providecommand \@url [1]{\endgroup\@href {#1}{\urlprefix }}%
\providecommand \urlprefix  [0]{URL }%
\providecommand \Eprint [0]{\href }%
\providecommand \doibase [0]{http://dx.doi.org/}%
\providecommand \selectlanguage [0]{\@gobble}%
\providecommand \bibinfo  [0]{\@secondoftwo}%
\providecommand \bibfield  [0]{\@secondoftwo}%
\providecommand \translation [1]{[#1]}%
\providecommand \BibitemOpen [0]{}%
\providecommand \bibitemStop [0]{}%
\providecommand \bibitemNoStop [0]{.\EOS\space}%
\providecommand \EOS [0]{\spacefactor3000\relax}%
\providecommand \BibitemShut  [1]{\csname bibitem#1\endcsname}%
\let\auto@bib@innerbib\@empty
\bibitem [{\citenamefont {Toader}\ \emph {et~al.}(2004)\citenamefont {Toader},
  \citenamefont {Covlea}, \citenamefont {Graham},\ and\ \citenamefont
  {Steen}}]{Toader2004}%
  \BibitemOpen
  \bibfield  {author} {\bibinfo {author} {\bibfnamefont {E.~I.}\ \bibnamefont
  {Toader}}, \bibinfo {author} {\bibfnamefont {V.}~\bibnamefont {Covlea}},
  \bibinfo {author} {\bibfnamefont {W.~G.}\ \bibnamefont {Graham}}, \ and\
  \bibinfo {author} {\bibfnamefont {P.~G.}\ \bibnamefont {Steen}},\ }\href
  {\doibase 10.1063/1.1637459} {\bibfield  {journal} {\bibinfo  {journal}
  {Review of Scientific Instruments}\ }\textbf {\bibinfo {volume} {75}},\
  \bibinfo {pages} {382} (\bibinfo {year} {2004})},\ \Eprint
  {http://arxiv.org/abs/https://doi.org/10.1063/1.1637459}
  {https://doi.org/10.1063/1.1637459} \BibitemShut {NoStop}%
\bibitem [{\citenamefont {Longmier}, \citenamefont {Baalrud},\ and\
  \citenamefont {Hershkowitz}(2006)}]{Longmier2006}%
  \BibitemOpen
  \bibfield  {author} {\bibinfo {author} {\bibfnamefont {B.}~\bibnamefont
  {Longmier}}, \bibinfo {author} {\bibfnamefont {S.}~\bibnamefont {Baalrud}}, \
  and\ \bibinfo {author} {\bibfnamefont {N.}~\bibnamefont {Hershkowitz}},\
  }\href {\doibase 10.1063/1.2393164} {\bibfield  {journal} {\bibinfo
  {journal} {Review of Scientific Instruments}\ }\textbf {\bibinfo {volume}
  {77}},\ \bibinfo {pages} {113504} (\bibinfo {year} {2006})},\ \Eprint
  {http://arxiv.org/abs/https://doi.org/10.1063/1.2393164}
  {https://doi.org/10.1063/1.2393164} \BibitemShut {NoStop}%
\bibitem [{\citenamefont {Robertson}(2013)}]{scott2013}%
  \BibitemOpen
  \bibfield  {author} {\bibinfo {author} {\bibfnamefont {S.}~\bibnamefont
  {Robertson}},\ }\href {http://stacks.iop.org/0741-3335/55/i=9/a=093001}
  {\bibfield  {journal} {\bibinfo  {journal} {Plasma Physics and Controlled
  Fusion}\ }\textbf {\bibinfo {volume} {55}},\ \bibinfo {pages} {093001}
  (\bibinfo {year} {2013})}\BibitemShut {NoStop}%
\bibitem [{\citenamefont {Scheiner}\ \emph {et~al.}(2016)\citenamefont
  {Scheiner}, \citenamefont {Baalrud}, \citenamefont {Hopkins}, \citenamefont
  {Yee},\ and\ \citenamefont {Barnat}}]{Brett2016}%
  \BibitemOpen
  \bibfield  {author} {\bibinfo {author} {\bibfnamefont {B.}~\bibnamefont
  {Scheiner}}, \bibinfo {author} {\bibfnamefont {S.~D.}\ \bibnamefont
  {Baalrud}}, \bibinfo {author} {\bibfnamefont {M.~M.}\ \bibnamefont
  {Hopkins}}, \bibinfo {author} {\bibfnamefont {B.~T.}\ \bibnamefont {Yee}}, \
  and\ \bibinfo {author} {\bibfnamefont {E.~V.}\ \bibnamefont {Barnat}},\
  }\href {\doibase 10.1063/1.4960382} {\bibfield  {journal} {\bibinfo
  {journal} {Physics of Plasmas}\ }\textbf {\bibinfo {volume} {23}},\ \bibinfo
  {pages} {083510} (\bibinfo {year} {2016})},\ \Eprint
  {http://arxiv.org/abs/https://doi.org/10.1063/1.4960382}
  {https://doi.org/10.1063/1.4960382} \BibitemShut {NoStop}%
\bibitem [{\citenamefont {Franklin}(2003)}]{Franklin2003}%
  \BibitemOpen
  \bibfield  {author} {\bibinfo {author} {\bibfnamefont {R.~N.}\ \bibnamefont
  {Franklin}},\ }\href {http://stacks.iop.org/0022-3727/36/i=22/a=R01}
  {\bibfield  {journal} {\bibinfo  {journal} {Journal of Physics D: Applied
  Physics}\ }\textbf {\bibinfo {volume} {36}},\ \bibinfo {pages} {R309}
  (\bibinfo {year} {2003})}\BibitemShut {NoStop}%
\bibitem [{\citenamefont {Stenzel}\ \emph {et~al.}(2011)\citenamefont
  {Stenzel}, \citenamefont {Gruenwald}, \citenamefont {Ionita},\ and\
  \citenamefont {Schrittwieser}}]{Stenzel2011}%
  \BibitemOpen
  \bibfield  {author} {\bibinfo {author} {\bibfnamefont {R.~L.}\ \bibnamefont
  {Stenzel}}, \bibinfo {author} {\bibfnamefont {J.}~\bibnamefont {Gruenwald}},
  \bibinfo {author} {\bibfnamefont {C.}~\bibnamefont {Ionita}}, \ and\ \bibinfo
  {author} {\bibfnamefont {R.}~\bibnamefont {Schrittwieser}},\ }\href {\doibase
  10.1063/1.3601858} {\bibfield  {journal} {\bibinfo  {journal} {Physics of
  Plasmas}\ }\textbf {\bibinfo {volume} {18}},\ \bibinfo {pages} {062112}
  (\bibinfo {year} {2011})},\ \Eprint
  {http://arxiv.org/abs/https://doi.org/10.1063/1.3601858}
  {https://doi.org/10.1063/1.3601858} \BibitemShut {NoStop}%
\bibitem [{\citenamefont {Hershkowitz}(2005)}]{Hershkowitz2005}%
  \BibitemOpen
  \bibfield  {author} {\bibinfo {author} {\bibfnamefont {N.}~\bibnamefont
  {Hershkowitz}},\ }\href {\doibase 10.1063/1.1887189} {\bibfield  {journal}
  {\bibinfo  {journal} {Physics of Plasmas}\ }\textbf {\bibinfo {volume}
  {12}},\ \bibinfo {pages} {055502} (\bibinfo {year} {2005})},\ \Eprint
  {http://arxiv.org/abs/https://doi.org/10.1063/1.1887189}
  {https://doi.org/10.1063/1.1887189} \BibitemShut {NoStop}%
\bibitem [{\citenamefont {Baalrud}, \citenamefont {Longmier},\ and\
  \citenamefont {Hershkowitz}(2009)}]{baalrud2009equilibrium}%
  \BibitemOpen
  \bibfield  {author} {\bibinfo {author} {\bibfnamefont {S.~D.}\ \bibnamefont
  {Baalrud}}, \bibinfo {author} {\bibfnamefont {B.}~\bibnamefont {Longmier}}, \
  and\ \bibinfo {author} {\bibfnamefont {N.}~\bibnamefont {Hershkowitz}},\
  }\href {\doibase 10.1088/0963-0252/18/3/035002} {\bibfield  {journal}
  {\bibinfo  {journal} {Plasma Sources Sci. Technol.}\ }\textbf {\bibinfo
  {volume} {18}},\ \bibinfo {pages} {035002} (\bibinfo {year}
  {2009})}\BibitemShut {NoStop}%
\bibitem [{\citenamefont {Mott-Smith}\ and\ \citenamefont
  {Langmuir}(1926)}]{Langmuir1926}%
  \BibitemOpen
  \bibfield  {author} {\bibinfo {author} {\bibfnamefont {H.~M.}\ \bibnamefont
  {Mott-Smith}}\ and\ \bibinfo {author} {\bibfnamefont {I.}~\bibnamefont
  {Langmuir}},\ }\href {\doibase 10.1103/PhysRev.28.727} {\bibfield  {journal}
  {\bibinfo  {journal} {Phys. Rev.}\ }\textbf {\bibinfo {volume} {28}},\
  \bibinfo {pages} {727} (\bibinfo {year} {1926})}\BibitemShut {NoStop}%
\bibitem [{\citenamefont {Song}, \citenamefont {D'Angelo},\ and\ \citenamefont
  {Merlino}(1991)}]{song1991anode}%
  \BibitemOpen
  \bibfield  {author} {\bibinfo {author} {\bibfnamefont {B.}~\bibnamefont
  {Song}}, \bibinfo {author} {\bibfnamefont {N.}~\bibnamefont {D'Angelo}}, \
  and\ \bibinfo {author} {\bibfnamefont {R.~L.}\ \bibnamefont {Merlino}},\
  }\href {\doibase 10.1088/0022-3727/24/10/012} {\bibfield  {journal} {\bibinfo
   {journal} {J. Phys. D: Appl. Phys.}\ }\textbf {\bibinfo {volume} {24}},\
  \bibinfo {pages} {1789} (\bibinfo {year} {1991})}\BibitemShut {NoStop}%
\bibitem [{\citenamefont {Chauhan}\ \emph {et~al.}(2016)\citenamefont
  {Chauhan}, \citenamefont {Ranjan}, \citenamefont {Bandyopadhyay},\ and\
  \citenamefont {Mukherjee}}]{chauhan2016droplet}%
  \BibitemOpen
  \bibfield  {author} {\bibinfo {author} {\bibfnamefont {S.}~\bibnamefont
  {Chauhan}}, \bibinfo {author} {\bibfnamefont {M.}~\bibnamefont {Ranjan}},
  \bibinfo {author} {\bibfnamefont {M.}~\bibnamefont {Bandyopadhyay}}, \ and\
  \bibinfo {author} {\bibfnamefont {S.}~\bibnamefont {Mukherjee}},\ }\href
  {\doibase 10.1063/1.4939029} {\bibfield  {journal} {\bibinfo  {journal}
  {Phys. Plasmas}\ }\textbf {\bibinfo {volume} {23}},\ \bibinfo {pages}
  {013502} (\bibinfo {year} {2016})}\BibitemShut {NoStop}%
\bibitem [{\citenamefont {Borgohain}\ and\ \citenamefont
  {Bailung}(2017)}]{Bailung2017}%
  \BibitemOpen
  \bibfield  {author} {\bibinfo {author} {\bibfnamefont {B.}~\bibnamefont
  {Borgohain}}\ and\ \bibinfo {author} {\bibfnamefont {H.}~\bibnamefont
  {Bailung}},\ }\href {\doibase 10.1063/1.5006133} {\bibfield  {journal}
  {\bibinfo  {journal} {Physics of Plasmas}\ }\textbf {\bibinfo {volume}
  {24}},\ \bibinfo {pages} {113512} (\bibinfo {year} {2017})}\BibitemShut
  {NoStop}%
\bibitem [{\citenamefont {Conde}, \citenamefont {Fontan},\ and\ \citenamefont
  {Lamb{\'a}s}(2006)}]{conde2006transition}%
  \BibitemOpen
  \bibfield  {author} {\bibinfo {author} {\bibfnamefont {L.}~\bibnamefont
  {Conde}}, \bibinfo {author} {\bibfnamefont {C.~F.}\ \bibnamefont {Fontan}}, \
  and\ \bibinfo {author} {\bibfnamefont {J.}~\bibnamefont {Lamb{\'a}s}},\
  }\href {\doibase 10.1063/1.2388265} {\bibfield  {journal} {\bibinfo
  {journal} {Phys. Plasmas}\ }\textbf {\bibinfo {volume} {13}},\ \bibinfo
  {pages} {113504} (\bibinfo {year} {2006})}\BibitemShut {NoStop}%
\bibitem [{\citenamefont {Scheiner}\ \emph {et~al.}(2015)\citenamefont
  {Scheiner}, \citenamefont {Baalrud}, \citenamefont {Yee}, \citenamefont
  {Hopkins},\ and\ \citenamefont {Barnat}}]{baalrud-scheiner-theory-ES}%
  \BibitemOpen
  \bibfield  {author} {\bibinfo {author} {\bibfnamefont {B.}~\bibnamefont
  {Scheiner}}, \bibinfo {author} {\bibfnamefont {S.~D.}\ \bibnamefont
  {Baalrud}}, \bibinfo {author} {\bibfnamefont {B.~T.}\ \bibnamefont {Yee}},
  \bibinfo {author} {\bibfnamefont {M.~M.}\ \bibnamefont {Hopkins}}, \ and\
  \bibinfo {author} {\bibfnamefont {E.~V.}\ \bibnamefont {Barnat}},\ }\href
  {\doibase 10.1063/1.4939024} {\bibfield  {journal} {\bibinfo  {journal}
  {Phys. Plasmas}\ }\textbf {\bibinfo {volume} {22}},\ \bibinfo {pages}
  {123520} (\bibinfo {year} {2015})}\BibitemShut {NoStop}%
\bibitem [{\citenamefont {Yee}\ \emph {et~al.}(2017)\citenamefont {Yee},
  \citenamefont {Scheiner}, \citenamefont {Baalrud}, \citenamefont {Barnat},\
  and\ \citenamefont {Hopkins}}]{EsheathYEE}%
  \BibitemOpen
  \bibfield  {author} {\bibinfo {author} {\bibfnamefont {B.~T.}\ \bibnamefont
  {Yee}}, \bibinfo {author} {\bibfnamefont {B.}~\bibnamefont {Scheiner}},
  \bibinfo {author} {\bibfnamefont {S.}~\bibnamefont {Baalrud}}, \bibinfo
  {author} {\bibfnamefont {E.}~\bibnamefont {Barnat}}, \ and\ \bibinfo {author}
  {\bibfnamefont {M.}~\bibnamefont {Hopkins}},\ }\href
  {http://iopscience.iop.org/10.1088/1361-6595/aa56d7} {\bibfield  {journal}
  {\bibinfo  {journal} {Plasma Sources Science and Technology}\ } (\bibinfo
  {year} {2017})}\BibitemShut {NoStop}%
\bibitem [{\citenamefont {Torv{\'e}n}\ and\ \citenamefont
  {Andersson}(1979)}]{torven1979dl}%
  \BibitemOpen
  \bibfield  {author} {\bibinfo {author} {\bibfnamefont {S.}~\bibnamefont
  {Torv{\'e}n}}\ and\ \bibinfo {author} {\bibfnamefont {D.}~\bibnamefont
  {Andersson}},\ }\href {\doibase 10.1088/0022-3727/12/5/012} {\bibfield
  {journal} {\bibinfo  {journal} {J. Phys. D: Appl. Phys.}\ }\textbf {\bibinfo
  {volume} {12}},\ \bibinfo {pages} {717} (\bibinfo {year} {1979})}\BibitemShut
  {NoStop}%
\bibitem [{\citenamefont {Tang}\ and\ \citenamefont
  {Chu}(2003)}]{tang2003anode}%
  \BibitemOpen
  \bibfield  {author} {\bibinfo {author} {\bibfnamefont {D.}~\bibnamefont
  {Tang}}\ and\ \bibinfo {author} {\bibfnamefont {P.~K.}\ \bibnamefont {Chu}},\
  }\href {\doibase 10.1063/1.1589592} {\bibfield  {journal} {\bibinfo
  {journal} {Journal of applied physics}\ }\textbf {\bibinfo {volume} {94}},\
  \bibinfo {pages} {1390} (\bibinfo {year} {2003})}\BibitemShut {NoStop}%
\bibitem [{\citenamefont {Andersson}(1981)}]{Andersson1981}%
  \BibitemOpen
  \bibfield  {author} {\bibinfo {author} {\bibfnamefont {D.}~\bibnamefont
  {Andersson}},\ }\href {http://stacks.iop.org/0022-3727/14/i=8/a=008}
  {\bibfield  {journal} {\bibinfo  {journal} {Journal of Physics D: Applied
  Physics}\ }\textbf {\bibinfo {volume} {14}},\ \bibinfo {pages} {1403}
  (\bibinfo {year} {1981})}\BibitemShut {NoStop}%
\bibitem [{\citenamefont {Cartier}\ and\ \citenamefont
  {Merlino}(1987)}]{steven1987}%
  \BibitemOpen
  \bibfield  {author} {\bibinfo {author} {\bibfnamefont {S.~L.}\ \bibnamefont
  {Cartier}}\ and\ \bibinfo {author} {\bibfnamefont {R.~L.}\ \bibnamefont
  {Merlino}},\ }\href {\doibase 10.1063/1.866093} {\bibfield  {journal}
  {\bibinfo  {journal} {The Physics of Fluids}\ }\textbf {\bibinfo {volume}
  {30}},\ \bibinfo {pages} {2549} (\bibinfo {year} {1987})},\ \Eprint
  {http://arxiv.org/abs/http://aip.scitation.org/doi/pdf/10.1063/1.866093}
  {http://aip.scitation.org/doi/pdf/10.1063/1.866093} \BibitemShut {NoStop}%
\bibitem [{\citenamefont {Holland}, \citenamefont {Fried},\ and\ \citenamefont
  {Morales}(1993)}]{Holland1993}%
  \BibitemOpen
  \bibfield  {author} {\bibinfo {author} {\bibfnamefont {D.~L.}\ \bibnamefont
  {Holland}}, \bibinfo {author} {\bibfnamefont {B.~D.}\ \bibnamefont {Fried}},
  \ and\ \bibinfo {author} {\bibfnamefont {G.~J.}\ \bibnamefont {Morales}},\
  }\href {\doibase 10.1063/1.860806} {\bibfield  {journal} {\bibinfo  {journal}
  {Physics of Fluids B: Plasma Physics}\ }\textbf {\bibinfo {volume} {5}},\
  \bibinfo {pages} {1723} (\bibinfo {year} {1993})},\ \Eprint
  {http://arxiv.org/abs/https://doi.org/10.1063/1.860806}
  {https://doi.org/10.1063/1.860806} \BibitemShut {NoStop}%
\bibitem [{\citenamefont {Chodura}(1982)}]{chodura1982}%
  \BibitemOpen
  \bibfield  {author} {\bibinfo {author} {\bibfnamefont {R.}~\bibnamefont
  {Chodura}},\ }\href {\doibase 10.1063/1.863955} {\bibfield  {journal}
  {\bibinfo  {journal} {The Physics of Fluids}\ }\textbf {\bibinfo {volume}
  {25}},\ \bibinfo {pages} {1628} (\bibinfo {year} {1982})},\ \Eprint
  {http://arxiv.org/abs/http://aip.scitation.org/doi/pdf/10.1063/1.863955}
  {http://aip.scitation.org/doi/pdf/10.1063/1.863955} \BibitemShut {NoStop}%
\bibitem [{\citenamefont {Gurlui}\ \emph {et~al.}(2005)\citenamefont {Gurlui},
  \citenamefont {Agop}, \citenamefont {Strat}, \citenamefont {Strat},\ and\
  \citenamefont {B{\=a}c{\=a}i{\c{t}}{\=a}}}]{gurlui2005dlpotential}%
  \BibitemOpen
  \bibfield  {author} {\bibinfo {author} {\bibfnamefont {S.}~\bibnamefont
  {Gurlui}}, \bibinfo {author} {\bibfnamefont {M.}~\bibnamefont {Agop}},
  \bibinfo {author} {\bibfnamefont {M.}~\bibnamefont {Strat}}, \bibinfo
  {author} {\bibfnamefont {G.}~\bibnamefont {Strat}}, \ and\ \bibinfo {author}
  {\bibfnamefont {S.}~\bibnamefont {B{\=a}c{\=a}i{\c{t}}{\=a}}},\ }\href
  {\doibase 10.1143/JJAP.44.3253} {\bibfield  {journal} {\bibinfo  {journal}
  {Jpn. J. Appl. Phys., Part 1}\ }\textbf {\bibinfo {volume} {44}},\ \bibinfo
  {pages} {3253} (\bibinfo {year} {2005})}\BibitemShut {NoStop}%
\bibitem [{\citenamefont {Baalrud}, \citenamefont {Hershkowitz},\ and\
  \citenamefont {Longmier}(2007)}]{baalrud-2007-global}%
  \BibitemOpen
  \bibfield  {author} {\bibinfo {author} {\bibfnamefont {S.~D.}\ \bibnamefont
  {Baalrud}}, \bibinfo {author} {\bibfnamefont {N.}~\bibnamefont
  {Hershkowitz}}, \ and\ \bibinfo {author} {\bibfnamefont {B.}~\bibnamefont
  {Longmier}},\ }\href {\doibase 10.1063/1.2722262} {\bibfield  {journal}
  {\bibinfo  {journal} {Phys. Plasmas}\ }\textbf {\bibinfo {volume} {14}},\
  \bibinfo {pages} {042109} (\bibinfo {year} {2007})}\BibitemShut {NoStop}%
\bibitem [{\citenamefont {Griem}(2005)}]{Griem}%
  \BibitemOpen
  \bibfield  {author} {\bibinfo {author} {\bibfnamefont {H.~R.}\ \bibnamefont
  {Griem}},\ }\href {\doibase 9780521619417} {\emph {\bibinfo {title}
  {Principles of Plasma Spectroscopy}}}\ (\bibinfo  {publisher} {Cambridge
  University},\ \bibinfo {year} {2005})\BibitemShut {NoStop}%
\bibitem [{\citenamefont {Mujawar}, \citenamefont {Karkari},\ and\
  \citenamefont {Turner}(2011)}]{karkari}%
  \BibitemOpen
  \bibfield  {author} {\bibinfo {author} {\bibfnamefont {M.~A.}\ \bibnamefont
  {Mujawar}}, \bibinfo {author} {\bibfnamefont {S.~K.}\ \bibnamefont
  {Karkari}}, \ and\ \bibinfo {author} {\bibfnamefont {M.~M.}\ \bibnamefont
  {Turner}},\ }\href {http://stacks.iop.org/0963-0252/20/i=1/a=015024}
  {\bibfield  {journal} {\bibinfo  {journal} {Plasma Sources Sci. Technol.}\
  }\textbf {\bibinfo {volume} {20}},\ \bibinfo {pages} {015024} (\bibinfo
  {year} {2011})}\BibitemShut {NoStop}%
\bibitem [{\citenamefont {Chaubey}\ \emph {et~al.}(2015)\citenamefont
  {Chaubey}, \citenamefont {Mukherjee}, \citenamefont {Sekar~Iyengar},\ and\
  \citenamefont {Sen}}]{chaubey2015sync}%
  \BibitemOpen
  \bibfield  {author} {\bibinfo {author} {\bibfnamefont {N.}~\bibnamefont
  {Chaubey}}, \bibinfo {author} {\bibfnamefont {S.}~\bibnamefont {Mukherjee}},
  \bibinfo {author} {\bibfnamefont {A.~N.}\ \bibnamefont {Sekar~Iyengar}}, \
  and\ \bibinfo {author} {\bibfnamefont {A.}~\bibnamefont {Sen}},\ }\href
  {\doibase 10.1063/1.4913227} {\bibfield  {journal} {\bibinfo  {journal}
  {Phys. Plasmas}\ }\textbf {\bibinfo {volume} {22}},\ \bibinfo {pages}
  {022312} (\bibinfo {year} {2015})}\BibitemShut {NoStop}%
\end{thebibliography}
%

\end{document}